\documentclass[aps,11pt,nofootinbib]{revtex4-1}
\pdfoutput=1

\usepackage{graphicx}
\usepackage[margin=1in]{geometry}
\usepackage{amsmath}
\usepackage{amssymb}
\usepackage{xcolor}
\usepackage[utf8]{inputenc}
\usepackage[all]{nowidow}

\usepackage{tensor}
\usepackage{braket}
\usepackage{float}
\usepackage[linktocpage=true,colorlinks=true,linkcolor=blue,citecolor=blue,urlcolor=blue]{hyperref}
\hypersetup{pdftitle={The power spectrum on small scales: Robust constraints and comparing PBH methodologies},pdfauthor={4 authors}}

\newcommand{\PBH}{\mathrm{PBH}}
\newcommand{\CDM}{\mathrm{CDM}}

\newcommand{\diffd}{\mathrm{d}} 
\newcommand{\ddv}[2]{\frac{\diffd #1}{\diffd #2}} 
\newcommand{\Msun}{\mathrm{M}_\odot}

\begin{document}

\author{Andrew D.~Gow$^1$}
\email{A.D.Gow@sussex.ac.uk}
\author{Christian T.~Byrnes$^1$}
\email{C.Byrnes@sussex.ac.uk}
\author{Philippa S.~Cole$^1$}
\email{P.Cole@sussex.ac.uk}
\author{Sam Young$^2$}
\email{syoung@mpa-garching.mpg.de}

\affiliation{\\1) Department of Physics and Astronomy, University of Sussex,\\Brighton BN1 9QH, United Kingdom\\} 

\affiliation{\\2) Max Planck Institute for Astrophysics,\\Karl-Schwarzschild-Strasse 1,
85748 Garching bei Muenchen, Germany}

\date{01/12/2020}

\title{The power spectrum on small scales: Robust constraints and comparing PBH methodologies}

\begin{abstract}
We compare primordial black hole (PBH) constraints on the power spectrum and mass distributions using the traditional Press Schechter formalism, peaks theory, and a recently developed version of peaks theory relevant to PBHs. We show that, provided the PBH formation criteria and the power spectrum smoothing are treated consistently, the constraints only vary by $\sim$~10\% between methods (a difference that will become increasingly important with better data). Our robust constraints from PBHs take into account the effects of critical collapse, the non-linear relation between $\zeta$ and $\delta$, and the shift from the PBH mass to the power spectrum peak scale. We show that these constraints are remarkably similar to the pulsar timing array (PTA) constraints impacting the black hole masses detected by LIGO and Virgo, but that the \mbox{$\mu$--distortion} constraints rule out supermassive black hole (SMBH) formation and potentially even the much lighter mass range of $\sim(1$--$100)\ \Msun$ that LIGO/Virgo probes.
\end{abstract}

\maketitle

\tableofcontents

\section{Introduction}

Primordial black holes (PBHs) could have formed in the early universe from the collapse of density perturbations \cite{Zel'dovich_1967,Hawking_1971,Carr_1974}. Although there are no confirmed detections of PBHs, there are tentative hints for their existence and in particular a lot of recent interest has focused on whether the Large Interferometer Gravitational-wave Observatory (LIGO) has detected PBHs \cite{Carr:2020xqk,Green:2020jor}. Assuming that PBHs formed from the collapse of large amplitude perturbations shortly after horizon entry during radiation domination, there is an approximate one-to-one relation between the scale at which the primordial power spectrum has a large amplitude peak and the mass of PBHs that form. See \cite{Green_2015,Sasaki_2018,Carr:2020gox}  for reviews.

In order for PBHs to form, the amplitude of the power spectrum must become orders of magnitude larger than the value of $2\times10^{-9}$ measured on large scales, e.g.~via observations of the cosmic microwave background (CMB) \cite{Akrami:2018odb}. Precisely how much larger it must become is a matter of active research, with significantly differing values being quoted in the literature, typically varying between $\mathcal{O}(10^{-3})$ and $\mathcal{O}(10^{-2})$ with values at the lower end quoted in e.g.~\cite{Dalianis:2018ymb,Sato-Polito:2019hws}.
$\mathcal{O}(10^{-1})$ values have also been considered in e.g.~\cite{Bringmann:2011ut}. Since the power spectrum amplitude is only logarithmically sensitive to the allowed energy density fraction of PBHs, this variation has little to do with the different PBH masses or constraints being considered and instead is primarily due to differences in the theoretical techniques being used to relate the power spectrum amplitude to the abundance of PBHs. Primordial non-Gaussianity also has an important impact on the required power spectrum amplitude, see e.g.~\cite{PinaAvelino:2005rm,Lyth:2012yp,Byrnes:2012yx,Young:2014oea,Young:2015cyn,Franciolini:2018vbk,Yoo:2019pma,Cai:2019_Gravitational}, but we will not consider that issue further in this paper. However, we do include an accurate approximation for the significant correction arising due to the non-linear relation between the density contrast and curvature perturbation, the importance of which has only recently been quantified \cite{Kawasaki:2019mbl,Young:2019yug,DeLuca:2019qsy,Yoo:2018kvb,Kalaja:2019uju}.

In this paper we make the first detailed study of how the PBH mass distribution differs when using Press Schechter or peaks theory as well as a recently developed treatment of peaks theory \cite{Young:2020xmk}, which solves a problem for PBHs related to the cloud-in-cloud problem. When a PBH forms, the final mass depends on both the amplitude and scale of the perturbation from which it forms \cite{Musco:2009_CC}, and the new treatment of peaks theory ensures that the amplitude of peaks are evaluated at the correct scale, giving the correct mass distribution and abundance. We also consider the sensitivity to the choice of the window function. We show that, provided that all quantities are calculated in a self-consistent way -- for example, the choice of window function must be reflected in the collapse threshold $\delta_c$ -- all techniques and window functions lead to quite consistent results whereby the uncertainty in the power spectrum amplitude is only of order $10\%$. This is a much smaller variation than \cite{Ando:2018qdb} found even due to just the choice of the window function alone, consistent with the corrections accounted for in Ref.~\cite{Young:2019osy}. We also note that, throughout this paper, we assume a fixed value for the collapse threshold of primordial perturbations. In reality, the exact value of the collapse threshold depends on the specific shape of each individual perturbation, and neglecting this gives an additional uncertainty of order a few percent \cite{Musco:2018rwt,Escriva:2020tak,Germani:2018jgr,Young:2019osy}.

The uncertainty in the initial conditions required to generate a required number of PBHs has important implications for relating observations of PBHs to observations of the associated enhanced amplitude of the primordial perturbations. This can be done, for example, via the observation of a stochastic background of gravitational waves measurable by pulsar timing arrays (PTAs) which measure frequencies corresponding to a horizon scale which could have formed the black holes observed by LIGO and Virgo. In general, understanding how to map from a PBH abundance to a power spectrum constraint is important for our understanding of the initial conditions of the universe and the constraints on models of inflation.

In the next section we introduce the calculation of the PBH mass distribution. In section \ref{sec:variability} we discuss how the result depends on the calculation technique and window function and we use these results to calculate robust constraints on the primordial power spectrum in section \ref{sec:Power_spectrum_constraints}, in particular showing that the pulsar timing array constraints are not inconsistent with the formation of LIGO mass PBHs. We conclude in section \ref{sec:conclusions}, and some technical details of the observational constraints and the non-linear mapping from the curvature perturbation to the density contrast are contained in the appendices.

\section{Obtaining the PBH mass distribution}
\label{sec:Obtaining_PBH_mass_distribution}
The procedure for obtaining the mass distribution from the power spectrum is similar for all three methods considered, and is based on connecting the PBH abundance $\Omega_\PBH$ to the mass fraction $\beta = \frac{\rho_\PBH(t_i)}{\rho(t_i)}$, where $\rho_\PBH$ is the mean energy density in PBHs, $\rho$ is the total background energy density, and $t_i$ is the time at which the PBHs form. This mass fraction is then related to the power spectrum. In every case, the PBH abundance is calculated from the mass fraction using
\begin{align}
\Omega_\PBH &= \int \diffd(\ln R)\ \frac{R_\text{eq}}{R}\beta(R), \label{eq:Omega_PBH}
\end{align}
where $R$ is the horizon scale at the time the PBH is forming, $R_\text{eq}$ is the horizon scale at matter-radiation equality and the ratio takes into account the relative growth of the PBH fraction during radiation domination. The form of $\beta(R)$ is different for each method, see eqs.~\eqref{eq:pressSchechter_beta}, \eqref{eq:tradPeaks_beta}, and \eqref{eq:modifiedPeak_beta}. The abundance is then related to the PBH mass function $f(m)$ through
\begin{align}
f(m) &= \frac{1}{\Omega_\CDM}\ddv{\Omega_\PBH}{(\ln m)},
\end{align}
which satisfies the normalisation condition
\begin{align}
\int \diffd(\ln m)\ f(m) &= f_\PBH = \frac{\Omega_\PBH}{\Omega_\CDM}.
\end{align}
This can then be related to the mass distribution $\psi(m)$ through
\begin{align}
\psi(m) &= \frac{1}{f_\PBH}\frac{f(m)}{m},
\end{align}
which is a PDF and hence satisfies the normalisation condition
\begin{align}
\int \diffd m\ \psi(m) &= 1.
\end{align}

The relation between $\beta(R)$ and the power spectrum then depends on the method used. In this paper, three methods are considered: a Press-Schechter-like calculation (PS), the traditional peaks theory method (TP) described in the classic BBKS paper \cite{Bardeen:1985tr}, and a modified peaks theory derived by Young and Musso (YM) \cite{Young:2020xmk}.

Recently, other variations of peaks theory have also been developed and applied to PBHs. Ref.~\cite{Yoo:2018kvb} proposed a method relating peaks in the curvature perturbation to peaks in the density field, with the caveat that the power spectrum is sufficiently narrow such that peaks of only one scale exist. Since we will here consider peaks in the power spectrum with a non-negligible width, we will not further consider the calculations presented in Ref.~\cite{Yoo:2018kvb}\footnote{A new paper, released at a similar date to this work, claims to have solved this issue \cite{Yoo:2020}, and applies peaks theory to $\Delta\zeta$, which is proportional to the (linear component of the) density contrast - meaning that it is similar to the peaks theory calculation considered here.}. Ref.~\cite{Germani:2019zez} proposed a similar method to Ref.~\cite{Young:2020xmk}, with 2 major differences. The first is that a top-hat window function is used instead of a Gaussian window function. The second difference is that, as well as extending peaks theory itself, Ref.~\cite{Germani:2019zez} simultaneously attempted to account for the non-linear relation between the density contrast\footnote{The authors made use of an analytic relationship between the linear and non-linear fields. The expression is valid only at the centre of spherically symmetric peaks when a top-hat window function is used, and it is not clear this is a valid equation to use to represent the entire field.}. However, as discussed further in appendix \ref{app:top-hat}, the top-hat window function has significant drawbacks making it unsuitable for use in this paper without an additional cut-off. We will therefore focus on comparing the YM calculation with previous ``traditional'' calculations, which is expected to be an improvement on traditional peaks theory by correctly accounting for the initial scale and amplitude of perturbations in calculating the final PBH mass.

Ref.~\cite{Kalaja:2019uju} discusses many complex points related to calculating the PBH abundance from the primordial power spectrum in detail. However, the calculation of the PBH abundance in Ref.~\cite{Kalaja:2019uju} makes numerous simplifying assumptions, using peaks theory in a method similarly to that presented in Ref.~\cite{Young:2014ana}. The calculations used in this paper improve upon this by accounting for the non-linearity of the density contrast, a non-zero width of the power spectrum, and the dependance of PBH mass upon both the scale and amplitude of the perturbations from which it formed.

In the Press-Schechter formalism, the mass fraction is related to a probability distribution in the compaction function $C$ by
\begin{align}
\beta(R) &= 2\int_{C_c}^{\infty} \diffd C\ \frac{m}{M_H}P(C), \label{eq:pressSchechter_beta}
\end{align}
where the compaction is a smoothed version of the density contrast $\delta$ (see eq.~\eqref{eq:Compaction}). The probability density function is given by
\begin{align}
P(C) &= \frac{1}{\sqrt{2\pi}\sigma_0(R)}\exp\left(-\frac{C^2}{2\sigma_0(R)^2}\right),
\end{align}
and the mass ratio $m/M_H$ takes into account the effect of critical collapse. In traditional peaks theory, the mass fraction is related to the number density of peaks, $n$, through
\begin{align}
\beta(R) &= (2\pi)^\frac{3}{2}R^3 \int_{C_c}^{\infty} \diffd C\ \frac{m}{M_H}\  n\left(\frac{C}{\sigma_0(R)}\right), \label{eq:tradPeaks_beta}
\end{align}
where the number density is a function of $\nu=C/\sigma_0$, given by \cite{Bardeen:1985tr}
\begin{align}
n(\nu) &= \frac{1}{3^{3/2}(2\pi)^2}\left(\frac{\sigma_1}{\sigma_0}\right)^3\nu^3\exp\left(-\frac{1}{2}\nu^2\right).
\end{align}

The modified peaks theory developed in Ref.~\cite{Young:2020xmk} also has $\beta$ related to $n$ in a similar way to eq.~\eqref{eq:tradPeaks_beta}, but with a factor of $R^4$ rather than $R^3$, i.e.
\begin{align}
\beta(R) &= (2\pi)^\frac{3}{2}R^4 \int_{C_c}^{\infty} \diffd C\ \frac{m}{M_H}\  n\left(\frac{C}{\sigma_0(R)}\right). \label{eq:modifiedPeak_beta}
\end{align}
This is required to counteract an extra inverse spatial dimension in the number density, given by
\begin{align}
n(\nu) &= \frac{16\sqrt{2}}{3^{3/2}\pi^{5/2}}\frac{\sigma_{RR}}{\sigma_2\sqrt{1-\gamma_{0,2}^2}R^7}\left(\frac{\sigma_0}{\sigma_1}\right)^3 \alpha\nu^4\exp\left(-\frac{1 + \frac{16\sigma_0^2}{R^4\sigma_2^2} - \frac{8\sigma_0\gamma_{0,2}}{R^2\sigma_2}}{1 - \gamma_{0,2}^2}\frac{\nu^2}{2}\right),
\end{align}
where $\gamma_{0,2}$ and $\alpha$ are related to the width parameters $\sigma_n(R)$ (see Ref.~\cite{Young:2020xmk} for more details). These width parameters relate the probability density (in Press-Schechter) or number density (in the peaks theories) to the power spectrum through the relation
\begin{align}
\sigma_n^2(R) &= \int_{0}^{\infty} \frac{\diffd k}{k}\ k^{2n}\,\mathcal{P}_{\delta_R}(k), \label{eq:sigma_n}
\end{align}
where $\mathcal{P}_{\delta_R}(k)$ is the compaction power spectrum, related to the power spectrum for $\zeta$ through
\begin{align}
\mathcal{P}_{\delta_R}(k) &= \frac{16}{81}(kR)^4 W^2(k,R) \mathcal{P}_\zeta(k).
\end{align}
$W(k,R)$ is a window function applied to the power spectrum. In this paper, two window functions are considered: a real-space top-hat\footnote{It should be noted that we have modified the top-hat window function to remove a ringing effect at large-$R$ (see appendix~\ref{app:WF-TH} for details).}, given in Fourier-space by
\begin{align}
W_\text{TH}(k,R) &= 3\frac{\sin(kR) - kR\cos(kR)}{(kR)^3}, \label{eq:WF-TH}
\end{align}
and a Gaussian window function modified by a factor of 2 in the exponent as suggested in Ref.~\cite{Young:2019osy},
\begin{align}
W_\text{G}(k,R) &= \exp\left(-\frac{(kR)^2}{4}\right). \label{eq:WF-G}
\end{align}
It should be noted that, in the case of the modified Gaussian window function, the compaction referred to by $C$ above is not technically the compaction, but is rather a ``compaction-like'' function. The compaction (or compaction-like function) is related to the PBH mass through the critical collapse equation,
\begin{align}
m &= KM_H(C-C_c)^\gamma, \label{eq:criticalScaling}
\end{align}
where $K$, $C_c$, and $\gamma$ are numerical factors that depend on the window function used to smooth the power spectrum, as well as the shape of the density perturbation \cite{Germani:2018jgr,Musco:2018rwt,Kalaja:2019uju}. The values $K\approx3.3$, $C_c\approx0.45$, and $\gamma\approx0.36$ (commonly referred to as the Musco criteria) were derived for the top-hat window function \cite{Niemeyer:1998_CC,Musco:2005_CC,Musco:2009_CC}, but are regularly used for other window functions. This has been highlighted in recent work, where different window functions cause a large deviation in the amplitude of power spectrum constraints, but this difference is not so significant if these numerical values are handled consistently for each window function \cite{Young:2019osy}. We will take the values stated in Ref.~\cite{Young:2020xmk}: $K=4$ and $C_c=0.55$ for the top-hat window function, and $K=10$ and $C_c=0.25$ for the modified Gaussian window function. For both window functions we take $\gamma=0.36$.

In this paper we will frequently consider a power spectrum with a lognormal peak, as a simple parametrisation of a peaked power spectrum with a position and width that can be easily tuned. The form is
\begin{align}\label{eq:lognormal}
{\cal P}_\zeta=A \frac{1}{\sqrt{2\pi}\Delta}\exp\left(-\frac{\ln^2(k/k_p)}{2\Delta^2}\right)
\end{align}
which has been appropriately normalised such that the constraint on $A$ becomes independent of $\Delta$ in the limit of a narrow peak, and it matches the delta function power spectrum $A\,\delta(\ln(k/k_p))$ in this limit. We show this later in table~\ref{tab:Width-dependence}. The integral of this power spectrum over $\ln k$ is $A$, independently of the value of $\Delta$. The width $\Delta$ is a free parameter, and we will normally choose two representative values for the width, $\Delta=0.3$ as a narrow peak which results in a PBH mass distribution not very different from that due to a delta-function power spectrum, and $\Delta=1$ as a broad peak which is roughly what one would expect if the inflaton field dynamics change over a time-scale of 1-efolding during inflation. We note that such a peak should not be extrapolated to values of $k$ very different in magnitude from $k_p$ (and of course the power spectrum needs to match the quasi scale-invariant spectrum observed on CMB scales), but in practice we have checked that both the power spectrum constraints and the PBH mass distribution do not depend on the shape of the peak when sufficiently far from the peak position (where the power spectrum amplitude is significantly smaller than the peak value). We are therefore not concerned (for the values of $\Delta$ we focus on) that a lognormal peak exhibits a growth steeper than $k^4$ on scales far from the peak, even though this is the approximate maximum growth rate of the power spectrum in canonical single-field inflation \cite{Byrnes:2018txb,Carrilho:2019oqg,Ozsoy:2019lyy}. A steeper growth can be achieved in e.g.~multifield inflation \cite{Palma:2020ejf,Fumagalli:2020adf}.

It is convenient to state the peak scale $k_p$ in terms of the horizon mass it corresponds to, using the relation derived by comparing the temperature of the radiation within the horizon mass with the temperature at matter-radiation equality in \cite{Nakama:2017_StochGW}
\begin{align}
M_H &= \frac{1}{\sqrt{2}} M_{\rm eq} \left(\frac{g_{\rm eq}}{g}\right)^{1/6}\left(\frac{k_{\rm eq}}{k}\right)^2\simeq17\left(\frac{g_*}{10.75}\right)^{-\frac{1}{6}}\left(\frac{k}{10^6\text{ Mpc}^{-1}}\right)^{-2}\Msun, \label{eq:MH(k)}
\end{align}
where $g_*$ is the number of relativistic degrees of freedom. We define the horizon mass at the peak of the power spectrum as
\begin{align}
M_{H,\mathcal{P}} &= M_H(k_p). \label{eq:MHP}
\end{align}

\section{Variability of the mass distribution}
\label{sec:variability}
\subsection{Effect of the calculation method and window function}
\label{ssec:Var_WF_Method}
Constraints on the PBH abundance can be used to place constraints on the amplitude of the primordial power spectrum. If the black holes in the LIGO merger events are considered to be primordial in origin, a fit of the masses and number of events can be used to constrain the PBH mass distribution, and hence the power spectrum. Recent studies have shown that, in this case, $f_\PBH$ would have to lie between $10^{-2}$ and $10^{-3}$, and would be closer to the lower of these two values \cite{Chen_2018,Raidal_2019,Gow:2019pok,DeLuca_2020}. 

See, however, recent papers \cite{Jedamzik:2020ypm,Young:2020scc,Jedamzik:2020omx,Clesse:2020ghq} discussing the effect of interactions between binary and single PBHs, which suggests that a much larger value for $f_\PBH$ is possible provided that PBH binaries are sufficiently disrupted by other PBHs. Ref.~\cite{Jedamzik:2020ypm} studied such 3-body interactions within extremely dense PBH clusters thought to form at high redshift when $f_\PBH\approx 1$, finding that the large majority of binaries in such clusters are expected to be disrupted, therefore not contributing to the merger rate observable today, implying that PBHs could make up the entirety of dark matter. Ref.~\cite{Young:2020scc} studied similar interactions within \emph{Milky Way}-type haloes, finding that the coalescence times can change significantly due to the interactions, especially when the PBH abundance is low. In addition, the effect of initial clustering of PBHs (due to primordial non-Gaussianity) on the merger rate was studied in Ref.~\cite{Young:2019gfc}, showing that this results in large uncertainties in the merger rate. Combined, these papers cast significant doubt on constraints on the PBH abundance coming from the observed merger rate.

However, in order to proceed with the comparison presented here, we will assume that the constraints are valid. Therefore, for each method and window function described above, we determine the power spectrum amplitude required to generate an $f_\PBH$ in the range $10^{-2} < f_\PBH < 10^{-3}$, chosen as $f_\PBH=2\times10^{-3}$. The resulting amplitudes are shown in table~\ref{tab:fPBH_amplitudes}. It should be noted that these amplitudes are defined for power spectrum peaks centred on the LIGO mass range, and would be significantly different on different scales. The full procedure for obtaining constraints on the power spectrum across all scales is described in section \ref{sec:Power_spectrum_constraints}.

\begin{table}[H]
\centering
\caption{Power spectrum amplitudes required to generate $f_\PBH=2\times10^{-3}$, with masses in the LIGO range. The two window functions are Gaussian (G) and Top-Hat (TH), and the three methods are Press-Schechter (PS), traditional Peaks theory (TP), and the modification to peaks theory calculated in Ref.~\cite{Young:2020xmk} (YM). The modified peaks theory cannot be applied in the case of a delta function peak, or with the top-hat window function, so these combinations are not shown.}\label{tab:fPBH_amplitudes}
\begin{tabular}{l|ccccc}
& \multicolumn{5}{c}{\textbf{Window Function, Method}} \\
\textbf{$\boldsymbol{\mathcal{P}}$ peak} & G, PS & G, TP & G, YM & TH, PS & TH, TP \\ \hline
Delta function & $3.21\times10^{-3}$ & $2.93\times10^{-3}$ & N/A & $3.47\times10^{-3}$ & $2.94\times10^{-3}$ \\
Lognormal ($\Delta=0.3$) & $4.14\times10^{-3}$ & $3.78\times10^{-3}$ & $3.55\times10^{-3}$ & $4.84\times10^{-3}$ & $4.13\times10^{-3}$ \\
Lognormal ($\Delta=1.0$) & $8.92\times10^{-3}$ & $8.14\times10^{-3}$ & $7.70\times10^{-3}$ & $1.11\times10^{-2}$ & $9.56\times10^{-3}$ \\
\end{tabular}
\end{table}

It can be seen that, when being careful with the combination of the window function and the corresponding critical collapse values, all the amplitudes are of the same order. When changing either the method or the window function while keeping the other fixed, the difference in the required amplitude is~$\lesssim20\%$. The biggest difference when taking both the window function and the calculation method into account is~$\sim32\%$. We note that the maximum value of the power spectrum does not vary nearly as much when $\Delta$ changes as suggested by table~\ref{tab:fPBH_amplitudes} due to our parametrisation of the power spectrum definition \eqref{eq:lognormal}. Choosing a different normalisation by leaving out the division by $\Delta$ would instead lead to a divergent value of the power spectrum amplitude in the limit $\Delta\rightarrow0$, instead of a value which matches the delta function power spectrum.

\begin{figure}[H]
\centering
\includegraphics[width=0.8\textwidth]{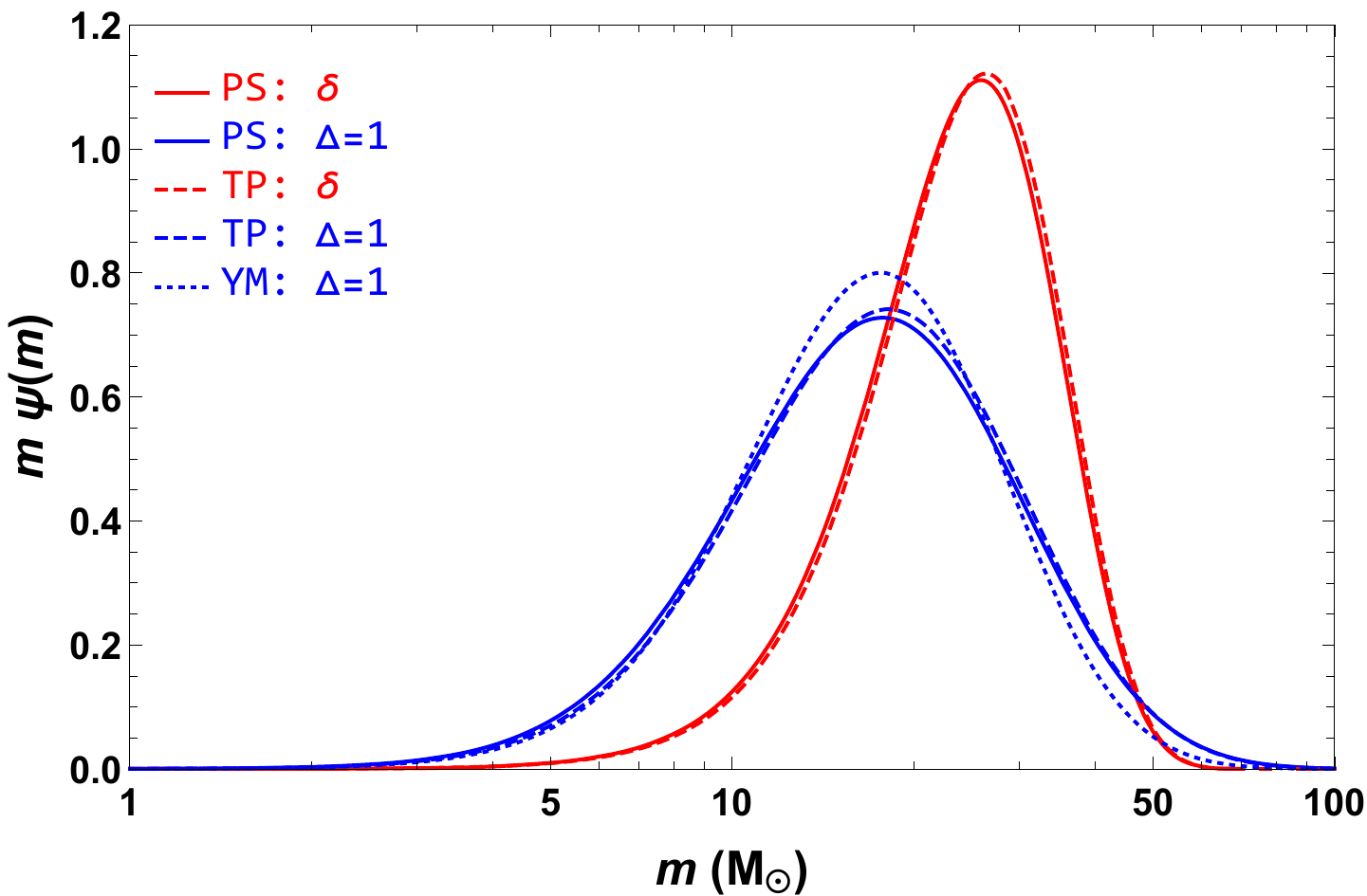}
\caption{Difference between PBH mass distributions calculated using different methods, while keeping the window function fixed. The Gaussian window function is used in every case. The red curves are for the delta function peak in the power spectrum, and the blue curves are for the lognormal peak with $\Delta=1$. The Press-Schechter (PS), traditional peaks (TP), and modified peaks (YM) methods are shown with solid, dashed, and dotted lines respectively. All lines have $f_\PBH=2\times10^{-3}$.}
\label{fig:psiPlot_M-All}
\end{figure}

We can also examine the amount of variability in the shapes of the mass distribution generated with different methods/window functions. The effect of changing the method is shown in fig.~\ref{fig:psiPlot_M-All}, for the Gaussian window function. The results for the top-hat case are similar. The mass distribution generated by a delta peak is shown in red, and the distribution for a lognormal peak with $\Delta=1$ in blue, with both peaks centered on $M_{H,\mathcal{P}}=4\ \Msun$ because this generates PBHs in the LIGO mass range. All the distributions are normalised to one, and correspond to $f_\PBH=2\times10^{-3}$. We find that the Press-Schechter (PS, solid) and peaks theory (TP, dashed) methods yield very similar results, while the modified peaks theory (YM, dotted) yields a marginally taller and narrower mass distribution.

\begin{figure}[H]
\centering
\includegraphics[width=0.8\textwidth]{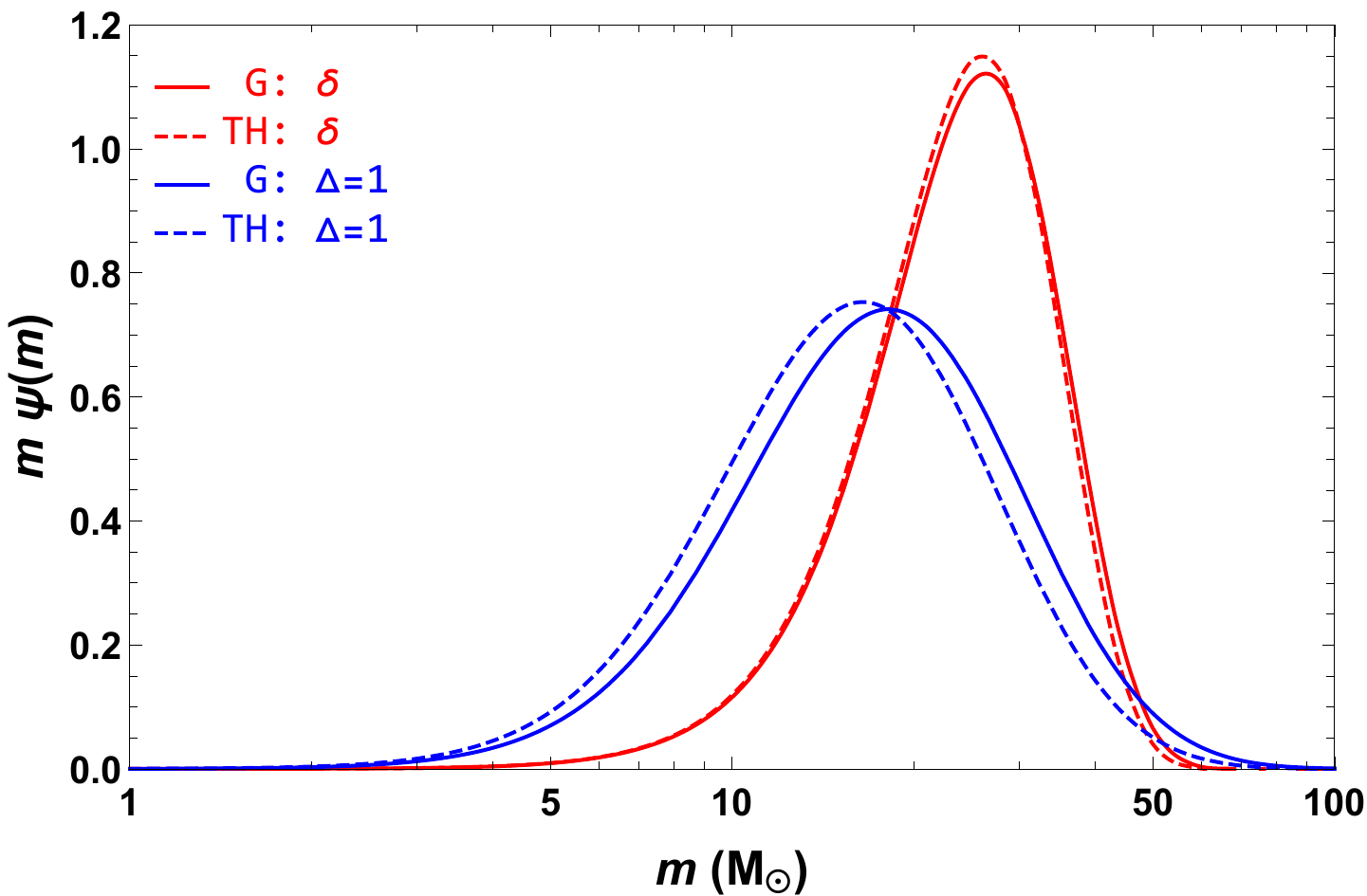}
\caption{Difference between PBH mass distributions calculated using different window functions, using the traditional peaks theory (TP) method. The red and blue curves correspond to a delta function power spectrum and a lognormal with $\Delta=1$ respectively. The solid and dashed lines are calculated using the Gaussian and top-hat window functions respectively. All lines have $f_\PBH=2\times10^{-3}$.}
\label{fig:psiPlot_WF-All}
\end{figure}

Figure~\ref{fig:psiPlot_WF-All} shows the effect of changing the window function, again for the delta function (red) and $\Delta=1$ lognormal (blue) cases, both with $M_{H,\mathcal{P}}=4\ \Msun$. All the distributions have been calculated using traditional peaks theory. The distributions calculated using the Gaussian and top-hat window functions are shown as solid and dashed lines respectively. The distributions from the two window functions are similar, but with a small shift in the peak position. Additionally, it can be seen from figs.~\ref{fig:psiPlot_M-All}~and~\ref{fig:psiPlot_WF-All} that there is a shift in the peak mass between the delta function power spectrum, and the $\Delta=1$ case. In the next section, we examine this shift in more detail for a range of power spectrum widths.

We have shown that the different calculation methods result in an $\mathcal{O}(10\%)$ shift in the required power spectrum amplitude, and a small difference in the shape and position of the mass distribution. We expect the BBKS peaks method (TP) to provide a more accurate result than the Press-Schechter (PS) case, since it can be viewed as a generalisation and collapses to the PS case under certain assumptions \cite{Wu:2020_Peak}, and that the modified version (YM) be better than TP, since it is a direct extension. Although the differences are small, they will become important in the future as experiments that can probe the PBH mass distribution become more accurate. For the remainder of this work, we will use the modified Gaussian window function in eq.~\eqref{eq:WF-G} and the traditional peaks theory (TP) method. This allows comparisons between other works that use the TP method and the results in this paper, which can then be compared between the different methods based on the differences highlighted here.

\subsection{Effect of the peak width $\Delta$}
\begin{figure}[H]
\centering
\includegraphics[width=0.8\textwidth]{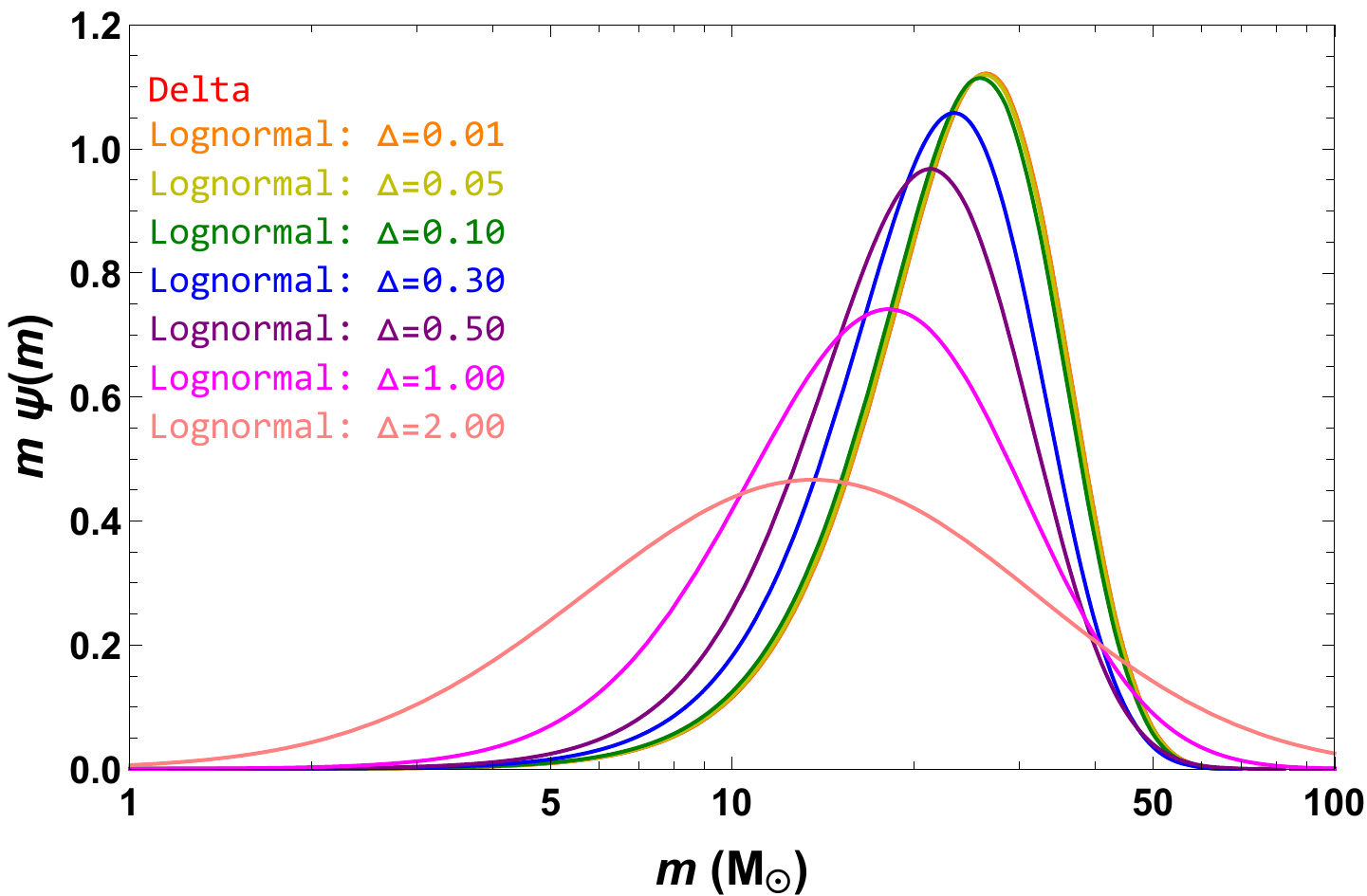}
\caption{Plot of the PBH mass distribution for different power spectrum peak widths $\Delta$. The peak position $k_p$ of the power spectrum is the same in every case, and corresponds to $M_{H,\mathcal{P}}=4\ \Msun$. All lines have $f_\PBH=2\times10^{-3}$. As $\Delta$ increases, the peak in the mass distribution shifts to smaller masses and spans a broader range of mass scales.}
\label{fig:psiPlot_Width-dependence}
\end{figure}

As shown in section \ref{ssec:Var_WF_Method}, the calculated mass distributions have a shift in the peak position which depends on the width of the power spectrum peak used. Additionally, we expect the width of the mass distribution to increase. We can demonstrate these effects by calculating the mass distributions for a range of values of $\Delta$ between zero (i.e.~a delta function peak) and two. The result of these calculations is shown in fig.~\ref{fig:psiPlot_Width-dependence}.

It is immediately apparent that, even for the unphysical choice of a delta function peak in the power spectrum, there is a minimum width in the mass distribution, associated with the critical collapse effect described in section~\ref{sec:Obtaining_PBH_mass_distribution}. It can also be seen that for very narrow peaks in the primordial power spectrum, the resulting mass distribution hardly varies until $\Delta\gtrsim0.1$. Beyond that point, the shift of the peak and the increased width become apparent. This means that whilst a monochromatic mass spectrum is unrealistic, studying a mass distribution with the minimum width due to critical collapse and a delta function power spectrum may be a good approximation to a physically realisable PBH mass distribution. The increasing width is also obvious, and can be quantified by fitting a lognormal mass distribution (the shape expected for PBHs arising from a smooth, symmetric peak) to data generated from the curves, and comparing the widths of these lognormals. The lognormal mass distribution is given by
\begin{align}
\psi(m) &= \frac{1}{\sqrt{2\pi}\sigma_\psi m}\exp\left(-\frac{\ln^2(m/m_c)}{2\sigma_\psi^2}\right),
\end{align}
where $m_c$ is the mean of the distribution and $\sigma_\psi$ is the width (note the subscript to avoid confusion with the $\sigma_n(R)$ parameters appearing in section~\ref{sec:Obtaining_PBH_mass_distribution}). The resulting lognormal parameters are shown in table~\ref{tab:Width-dependence}, and show that, as expected, the width of the calculated mass distribution increases with the peak width, as well as the amplitude required to keep $f_\PBH$ fixed. This minimum width appears to be much larger than is required in order for PBH decay to result in a sufficiently rapid transition from an early matter dominated era (caused by low mass PBHs) to radiation domination to generate an observable stochastic background of gravitational waves  \cite{Inomata:2020lmk}.

A noteworthy point here is that the typical mass of a PBH is actually significantly larger than the horizon mass corresponding to the scale at which the power spectrum peaks, $m_c/M_{H,\mathcal{P}}>1$. At first glance, this statement may seem to be in disagreement with previous works where the expected PBH mass has been shown to be smaller than the horizon mass at re-entry. Physically, this apparent discrepancy is due to the fact that, if there is a narrow peak in the $\zeta$ power spectrum at a scale $k_p$, the resultant perturbations will, on average, have a significantly larger characteristic scale $r_m$. In the calculation presented here, this manifests itself in the fact that the variance $\sigma_0^2(R)$ peaks at a larger value of $R$ than that corresponding to the scale $k_p$ (as calculated in Ref.~\cite{Germani:2018jgr} for example). Thus, the final mass of PBHs is smaller than the horizon mass corresponding to $r_m$, but larger than the horizon mass corresponding to $k_p$. The important conclusion drawn from this is that constraints on the PBH abundance for a given mass of PBH correspond to constraints on the primordial power spectrum at a larger value of $k$ than have previously been calculated.

\begin{table}[H]
\centering
\caption{Comparison of the amplitude required to generate $f_\PBH=2\times10^{-3}$, the ratio of the mean PBH mass $m_c$ to the power spectrum peak mass $M_{H,\mathcal{P}}$, and the mass distribution width $\sigma_\psi$ for different power spectrum peak widths $\Delta$.}
\label{tab:Width-dependence}
\resizebox{\textwidth}{!}{\begin{tabular}{c|c|c|c}
\textbf{$\boldsymbol{\mathcal{P}}$ peak width $\boldsymbol{\Delta}$} & \textbf{Required amplitude} $\boldsymbol{A}$ & \textbf{Mean PBH mass $\boldsymbol{m_c}$/peak mass $\boldsymbol{M_{H,\mathcal{P}}}$} & \textbf{Mass function width $\boldsymbol{\sigma_\psi}$} \\ \hline
0 (Delta) & $2.93\times10^{-3}$ & 6.21 & 0.374 \\
0.01 & $2.94\times10^{-3}$ & 6.21 & 0.374 \\
0.05 & $2.96\times10^{-3}$ & 6.17 & 0.375 \\
0.10 & $3.04\times10^{-3}$ & 6.09 & 0.377 \\
0.30 & $3.78\times10^{-3}$ & 5.52 & 0.395 \\
0.50 & $4.89\times10^{-3}$ & 5.07 & 0.430 \\
1.00 & $8.14\times10^{-3}$ & 4.39 & 0.553 \\ 
2.00 & $1.51\times10^{-2}$ & 3.35 & 0.864 \\ \hline
\end{tabular}}
\end{table}

Now we have a clear picture of how the different method and window function choices affect the mass distribution $\psi$ and the amplitude required to generate a fixed $f_\PBH$, we can calculate the constraints on the power spectrum from PBHs, being careful about the consistency of our window function and critical collapse choices. We show the procedure for obtaining these constraints, and the final constraint plots, in the next section.

\section{The constraints on the power spectrum}
\label{sec:Power_spectrum_constraints}
\subsection{Relevant constraints and how they are calculated}
Whilst calculating the PBH abundance with different methods has a huge effect on the calculated abundance and mass distribution, we have shown that the resultant uncertainty in constraints on the power spectrum is relatively small. We will now consider how observational limits on the PBH abundance, as well as a swathe of other observational probes, constrain the amplitude of the primordial power spectrum. The key additional constraints on small scales come from cosmic $\mu$-distortions \cite{Chluba:2012_Probing} and a stochastic background of gravitational waves, which could be generated with a large amplitude due to the non-linear coupling between the scalar and tensor perturbations around the time of horizon entry \cite{Tomita:1967_Non-Linear,Ananda_2007}. The calculation of many of these constraints follows closely the procedure presented in Ref.~\cite{Byrnes:2018txb}, and we therefore relegate the details to appendix \ref{app:observations}. However, we describe the constraints from PBHs in detail here, and we also highlight that constraints from PTAs have been updated to use the improved analysis of the NANOGrav 11 year data set \cite{Arzoumanian_2018}. There are additional small-scale constraints on the power spectrum, including for example those from y-distortions \cite{Chluba:2015bqa,Lucca:2019rxf}, 21cm observations \cite{Gong:2017sie,Bernal:2017nec,Mena:2019nhm,Munoz:2019hjh,Cole:2019zhu} and the non-detection of ultra-compact minihaloes  \cite{Bringmann:2011ut,Gosenca:2017ybi,Delos:2017thv,Delos:2018ueo,Furugori:2020jqn}. We do not display the former because the combination of CMB constraints and \mbox{$\mu$-distortion} constraints are more competitive on commensurate scales, and we do not display either of the latter because they depend on the dark matter model. Big Bang nucleosynthesis constraints are discussed in e.g.~\cite{Jeong:2014gna,Nakama:2014vla,Inomata:2016uip}.

\subsection{Constraints due to the gravitational wave background}\label{sec:GW}

Large amplitude scalar perturbations re-entering the horizon after inflation induce gravitational waves as a second-order effect. These contribute to the stochastic gravitational background, which pulsar timing arrays (PTAs) are trying to detect and/or constrain by looking for global changes in the time of arrival of pulses from a population of millisecond pulsars over a period of $\mathcal{O}(10)$ years. Details of the calculation of the GW power spectrum are contained in appendix \ref{app:GW}.

Translating this power spectrum to $\Omega_{\rm GW}h^2$ with eq.~\eqref{eq:om}, we can then compare the predicted signal with PTA constraints from the NANOGrav 11 year data set.\footnote{During the refereeing process, NANOGrav released their 12.5 year dataset \cite{2020arXiv200904496A} which showed \textit{possible} evidence for a signal due to a stochastic gravitational wave background. This is unconfirmed, but understanding the origin of this signal could have significant implications for the induced gravitational wave constraints discussed in this work.} We choose this data set because the new analysis takes errors in the modelling of the solar system ephemeris into account. This can have a large effect on the constraints which will need to be factored into the previous NANOGrav 9 year constraints \cite{Arzoumanian_2016}, as well as those from other arrays such as the European Pulsar Timing Array (EPTA) \cite{Lentati_2015} which have previously been used to constrain the primordial power spectrum with induced gravitational waves. Those constraints should now be revised upwards, but the analysis would need to be redone in each case to quantify by exactly how much. Based on the current analyses, the constraint on the characteristic strain $h_c$ improves by a factor of a third at the frequency of the tightest constraint, and improves by up to a factor of 5 at the highest frequencies between the 9 year and 11 year datasets. The resulting improvement on the primordial power spectrum constraint is shown in figure \ref{fig:ng119} of appendix \ref{app:ng}. Since the NANOGrav data set has pulsar timing data for 11 years of observations, it does not extend to quite as large scales as does the EPTA data, which is from 18 years of observations. This means that our constraints do not span as wide a range of scales (and hence PBH masses) as previous constraints in the literature show, but the constraints we do show are more robust to errors in solar system ephemeris modelling. We also avoid confusion over different analyses from different data sets, and are able to use the free spectrum constraints on $\Omega_{\rm GW}h^2$ consistently throughout.

These constraints (taken from the bottom panel of fig.~3 in Ref.~\cite{Arzoumanian_2018}) are the 2-$\sigma$ constraints derived as a function of frequency so as to represent the sensitivity to monochromatic signals. This means that we will construct our constraints based on finding the limiting amplitude of the lognormal power spectrum to which the NANOGrav constraints would be sensitive. One could do a more sophisticated analysis, taking into account the fact that confidence in a detection would become even stronger if there are also weaker detections of a given signal on larger or smaller frequencies than where the strongest detection would come. We choose to just show the 2-$\sigma$ constraints for clarity. We convert from frequency to scale with $k=2\pi c/f$ and then find the minimum value of $A$ for which $\Omega_{\rm GW, NG}h^2=\Omega_{\rm GW, signal}h^2$, i.e.
\begin{align}
    A_{\rm constraint}={\rm Min}\left(\sqrt{\frac{\Omega_{\rm GW, NG}(k)h^2}{\Omega_{\rm GW, signal}(k,k_p)h^2}}\right)
\end{align}
for each $k_p$. The minimum value of $A$ for each $k_p$ is found by scanning over all values of $k$ for which NANOGrav has sensitivity. We plot the results in figs.~\ref{fig:combinedConstraints_current}~and~\ref{fig:combinedConstraints_future} for $\Delta=0.3$ and $\Delta=1$, where again to be clear, the constraint on $\mathcal{P}_\mathcal{R}$ at a given $k$ represents the maximum amplitude $A$ for a lognormal power spectrum centred at $k=k_p$ such that the induced second-order gravitational waves would not be in conflict with the PTA constraints from the NANOGrav 11 year data set.

\subsection{Constraints from PBHs}
\begin{figure}[H]
\centering
\includegraphics[width=0.5\textwidth]{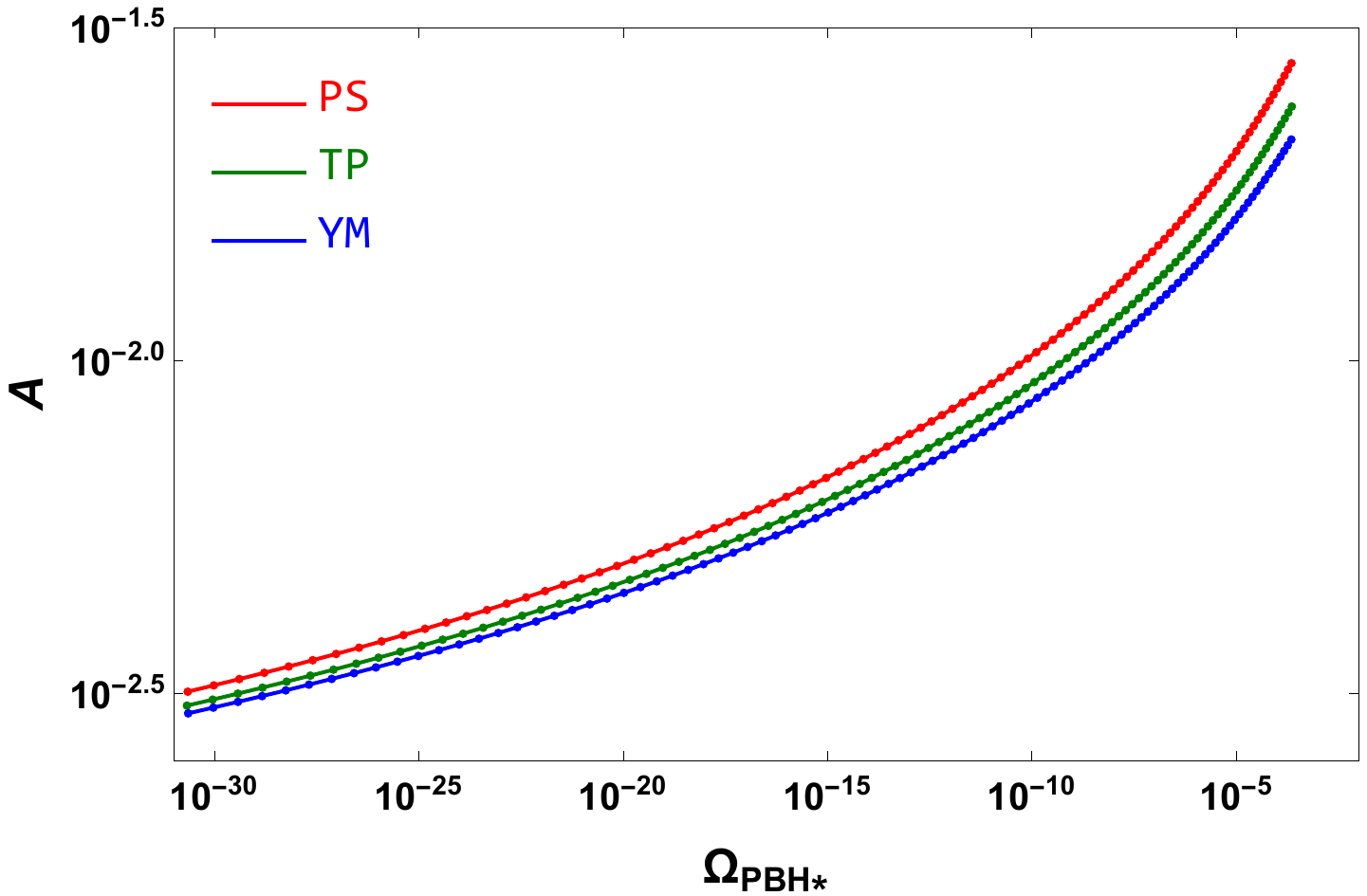}\includegraphics[width=0.5\textwidth]{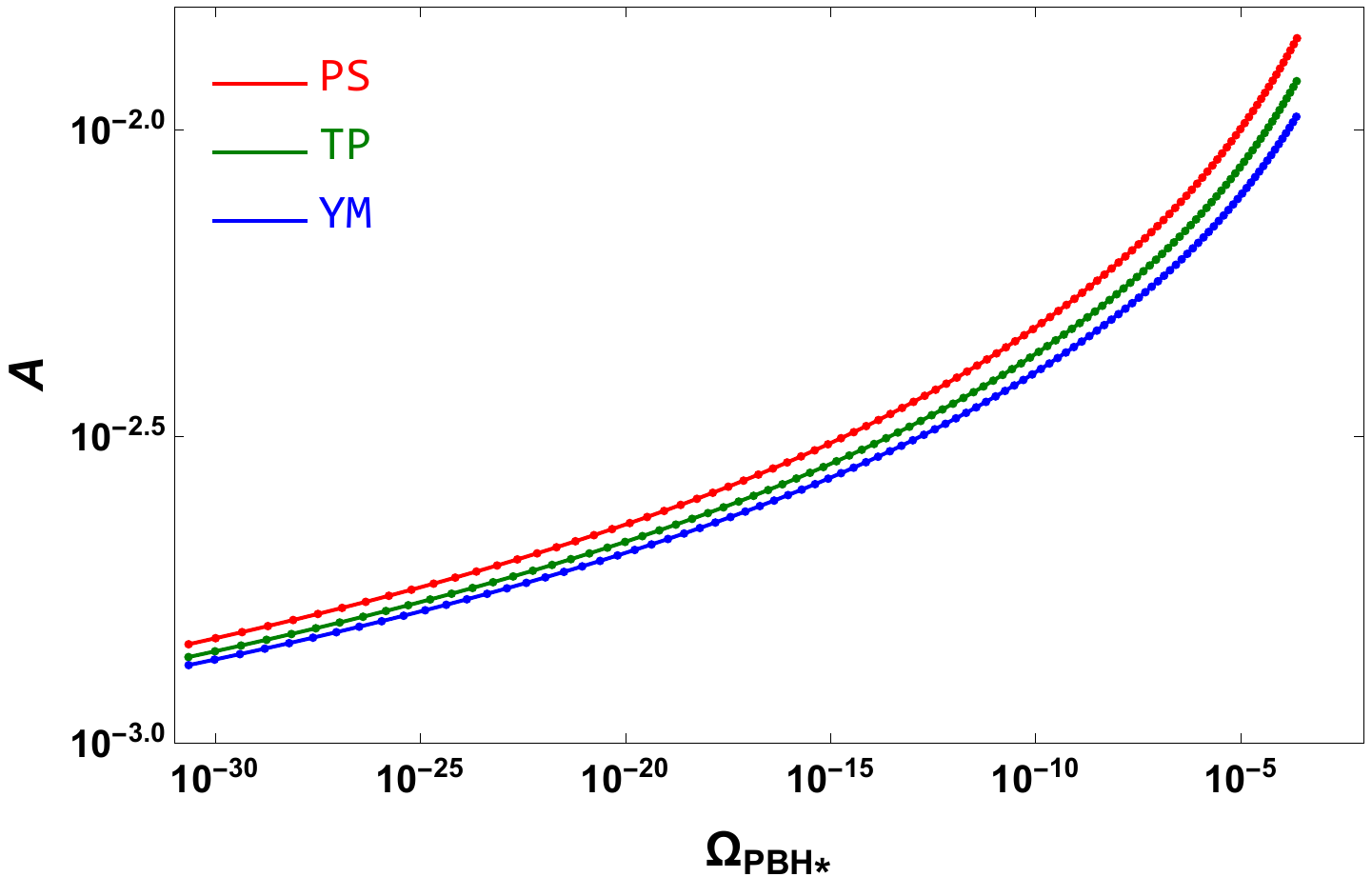}
\caption{Relation between power spectrum amplitude $A$ and $\Omega_{\PBH*}$ for the three methods. The power spectrum peak widths are $\Delta=1$ (left) and $\Delta=0.3$ (right). All lines use the Gaussian window function.}
\label{fig:AvsOmegaPBH}
\end{figure}

Constraints on primordial black holes are normally presented in terms of either $f_\PBH$ or the mass fraction $\beta$, so a method is required to relate these to the power spectrum amplitude. A relation between $f_\PBH$ (or equivalently $\Omega_\PBH$) is complicated by the fact that the redshifting factor in eq.~\eqref{eq:Omega_PBH} means that the required amplitude to generate a fixed $f_\PBH$ varies with the peak positions (as demonstrated in sec \ref{ssec:Var_WF_Method}). In general, the best way to overcome this would be to produce a relation for $A$ as a function of both $f_\PBH$ and the relevant mass scale. However, this is computationally expensive, and so a simplified approach is necessary. We can find an approximation by relating the power spectrum amplitude to a parameter that does not vary with the peak position, which we achieve by modifying eq.~\eqref{eq:Omega_PBH}, adjusting the redshift factor by introducing a new scale $R_*$, such that
\begin{align}
\Omega_{\PBH*} &= \int \diffd(\ln R)\ \frac{R_*}{R}\beta(R).
\end{align}
If $R_*$ is chosen to be close enough to the peak scale in the power spectrum, then the relation between this quantity and the power spectrum amplitude will be independent of the peak position. This quantity cannot be treated exactly as the abundance, because the abundance is calculated in the super-horizon regime before PBHs form, whereas this is at some later time, corresponding to when the horizon scale is $R_*$. This quantity can be related to the constraints for PBHs using
\begin{align}
\Omega_\PBH &= \frac{R_\text{eq}}{R*}\Omega_{\PBH*}. \label{eq:OmegaPBHStar_vs_OmegaPBH}
\end{align}
The relation between the power spectrum amplitude and $\Omega_{\PBH*}$ for all three methods is shown in fig.~\ref{fig:AvsOmegaPBH} for the $\Delta=1$ (left) and $\Delta=0.3$ (right) cases. The modified Gaussian window function is used in every case. It can be seen that there is a shift in the amplitude required between the methods, as was observed earlier. However, comparing the scale of changes to the power spectrum amplitude between the CMB value of $10^{-9}$ and these values, the differences are unimportant. For the constraint plots shown in figs.~\ref{fig:combinedConstraints_current}~and~\ref{fig:combinedConstraints_future}, the traditional peaks theory method (TP) is used.

To obtain constraints on the power spectrum, $\Omega_{\PBH*}$ must be related to constraints on either $\beta$ or $f_\PBH$. We will us the PBH constraints stated in Ref.~\cite{Carr:2020gox} for $\beta'$, which is a version of the mass fraction $\beta$ with common parameters normalised out. These constraints are calculated assuming that all the PBHs form at the same time (or equivalently, the same scale $R$), but it is possible to relate the constraints to $\Omega_{\PBH*}$, and hence determine the constraints on the amplitude for the calculation used throughout this paper, where PBHs form over a range of different scales. We obtain this relation from eqs.~(6) and (8) from \cite{Carr:2020gox} (reproduced here for clarity):
\begin{align}
\beta(m_c) = 7.06\times10^{-18}\gamma^{-1/2}&\left(\frac{h}{0.67}\right)^2\left(\frac{g_{*,i}}{106.75}\right)^{1/2}\left(\frac{m_c}{10^{15} \text{ g}}\right)^{1/2}\Omega_\PBH(m_c), \label{eq:betaCarr} \\
\beta'(m_c) = \gamma^{1/2}&\left(\frac{h}{0.67}\right)^{-2}\left(\frac{g_{*,i}}{106.75}\right)^{-1/2}\beta(m_c), \label{eq:betaPCarr}
\end{align}
where the monochromatic PBH mass $M$ in Ref.~\cite{Carr:2020gox} has been substituted for the mean lognormal mass $m_c$ (the constraints do not change significantly when considering a reasonably narrow PBH mass distribution \cite{Carr_2017,Bellomo_2018}). It can immediately be seen that, combining eqs.~\eqref{eq:betaCarr} and \eqref{eq:betaPCarr},
\begin{align}
\beta'(m_c) &= 7.06\times10^{-18}\left(\frac{m_c}{10^{15} \text{ g}}\right)^{1/2}\Omega_\PBH(m_c).
\end{align}
Since solar mass PBHs are of special interest, it is sensible to rescale the mass fraction to be in terms of solar masses, giving
\begin{align}
\beta'(m_c) &= 7.06\times10^{-18}\left(2\times10^{18}\frac{m_c}{\Msun}\right)^{1/2}\Omega_\PBH(m_c) \\
&= 10^{-8}\left(\frac{m_c}{\Msun}\right)^{1/2}\Omega_\PBH(m_c).
\end{align}
Inverting this relation gives $\Omega_\PBH$ as a function of $m_c$ in solar masses,
\begin{align}
\Omega_\PBH(m_c) = 10^8\left(\frac{m_c}{\Msun}\right)^{-1/2}\beta'(m_c).
\end{align}
We can then be relate this to the quantity $\Omega_{\PBH*}$ using eq.~\eqref{eq:OmegaPBHStar_vs_OmegaPBH} to give
\begin{align}
\Omega_{\PBH*}(m_c) &= 10^8\frac{R_*}{R_\text{eq}}\left(\frac{m_c}{\Msun}\right)^{-1/2}\beta'(m_c). \\
\intertext{For convenience, we have chosen $R_*$ such that the corresponding mass scale $M_*$ is approximately $m_c$. Therefore,}
\Omega_{\PBH*}(m_c) &= 10^8\left(\frac{m_c}{M_\text{eq}}\right)^{1/2}\left(\frac{m_c}{\Msun}\right)^{-1/2}\beta'(m_c) \\
&= 10^8\left(\frac{M_\text{eq}}{\Msun}\right)^{-1/2}\beta'(m_c).
\end{align}
Substituting in the value of the horizon mass at matter-radiation equality, $M_\text{eq}=2.8\times10^{17}\ \Msun$, the relation becomes
\begin{align}
\Omega_{\PBH*}(m_c) &\approx 0.2\ \beta'(m_c).
\end{align}

Recent papers \cite{Young:2019yug,Yoo:2018kvb,Kawasaki:2019mbl,Kalaja:2019uju,DeLuca:2019qsy} have discussed the effect of the non-linear relation between the curvature perturbation $\zeta$ and the density contrast $\delta$ on the PBH abundance. The point is that, even if the level of primordial non-Gaussianity of $\zeta$ is taken to be zero, $\delta$ will not have a Gaussian distribution, and subsequently nor will the compaction. The non-linearity is difficult to account for, especially if window functions other than a top-hat are considered. This is discussed in some detail in appendix \ref{app:nonLinearity}, with the conclusion that constraints on the power spectrum will be approximately 1.98 times weaker once the non-linearity is included in the calculation. We include this factor in the PBH lines in figs.~\ref{fig:combinedConstraints_current}~and~\ref{fig:combinedConstraints_future}.

By applying the method described in this section, we are taking into account the effects of critical collapse (making sure it is treated consistently with the choice of window function), the shift between the PBH mass and the peak scale $k_p$, and the non-linear relation between $\zeta$ and $\delta$. This is the first time that all of these effects have been captured simultaneously.

\subsection{Summarising all the constraints}
\begin{figure}[H]
\centering
\includegraphics[width=0.9\textwidth]{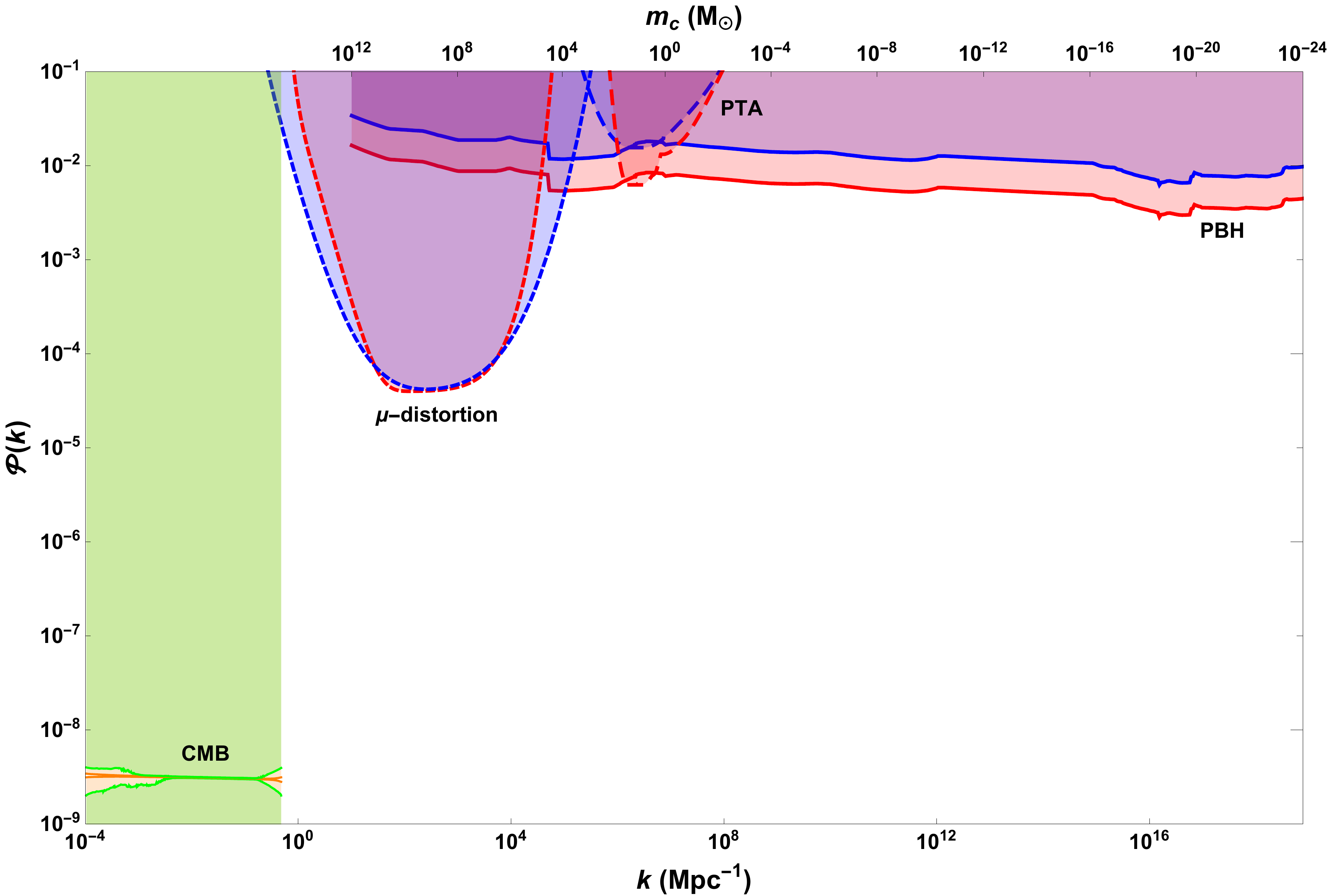}
\caption{Plot of the constraints on the power spectrum amplitude from PBH, PTA, and $\mu$-distortion sources, as well as the measured one and three-sigma constraints from the CMB. The constraints for $\Delta=0.3$ (which are tighter for the PBH constraints, and narrower for the other constraints) are shown in red, and the constraints for $\Delta=1$ are shown in blue. The PBH, PTA, and $\mu$-distortion constraints are shown with solid, long-dashed, and short-dashed lines respectively.}
\label{fig:combinedConstraints_current}
\end{figure}

In fig.~\ref{fig:combinedConstraints_current} we put together the key observational constraints to show the principal current constraints on the primordial power spectrum. The power spectrum has been accurately measured on large scales whilst PBHs constrain -- albeit weakly -- a far larger range of scales. We do not show PBH constraints on masses close to matter-radiation equality because we always assume PBHs form during radiation domination, and the smallest scale constrained corresponds to a PBH with $m_c\sim10^{-24}\ \Msun$, which evaporates around the time of Big Bang nucleosynthesis.

By coincidence the PTA measurements constrain the power spectrum amplitude to almost the same amplitude as the non-detection of PBHs, meaning that there is a potential tension between the PTA bounds and any claim that LIGO detected PBHs (see fig.~\ref{fig:combinedConstraints_current}). This has been studied by various groups \cite{Inomata:2016rbd,Byrnes:2018txb,Inomata:2018epa,Clesse:2018ogk,Lu:2019sti,Chen:2019xse,Wang:2019kaf,Yuan:2019udt,Cai:2019elf,Dalianis:2020_Exploring}, with no consensus reached on how severe the tension is. The impact of the PBH density profile was studied in depth in Ref.~\cite{Kalaja:2019uju} but the PTA constraint was not varied to reflect changes in the shape of the primordial power spectrum. For example \cite{Chen:2019xse} claim that $f_{\rm PBH}<10^{-6}$ over a significant range of PBH masses and the power spectrum constraint plots in Ref.~\cite{Byrnes:2018txb} appear to show a significant tension. By making a careful study of the power spectrum amplitude required to generate PBHs, including the important reduction in the PBH constraining power due to the non-linear relation between $\zeta$ and $\delta$, and using improved NANOGrav constraints, we have shown that there is no significant tension between generating LIGO mass PBHs and the PTA constraints.

We note that the slight overlap between the PBH and PTA constraint lines is not significant given the remaining ${\mathcal O}(10\%)$ uncertainty in the amplitude of the PBH constraint, and that there should also be about an ${\mathcal O}(10\%)$ reduction in the PBH line at about the $\Msun$ scale caused by the reduction in the equation-of-state parameter during the QCD transition. See \cite{Byrnes_2018} for further discussion, and \cite{Carr:2019kxo} for extensions to other masses where there is a smaller reduction in pressure within standard model physics. A study of non-standard expansion histories (such as an early matter dominated epoch) are beyond the scope of this paper \cite{Allahverdi:2020bys}. Nonetheless, because the PBH amplitude only depends very weakly on the value of $f_{\rm PBH}$ it is clear that the PTA collaborations should be very close to detecting a stochastic gravitational wave background even if only one of the compact objects which LIGO has detected was a PBH, for example the secondary mass object in the recently detected event  which falls into the mass gap between neutron stars and astrophysical black holes \cite{Abbott:2020khf}. It seems plausible that the associated stochastic background could be detectable with current PTA data if a dedicated search was made by using specific GW templates generated by power spectra that cause LIGO mass PBHs to form.

The cosmic $\mu$--distortion places an upper limit on the maximum PBH mass which can be generated by the collapse of large amplitude perturbations shortly after horizon reentry. The maximum mass decreases as the power spectrum width $\Delta$ increases, but even for a narrow peak with $\Delta=0.3$ the initial PBH mass cannot be much greater than $10^4\ \Msun$, which is much smaller than the supermassive BHs seen in the centre of most galaxies even at high redshift, with masses $10^6$--$10^9\ \Msun$, whose origin remains a mystery. However, such large PBHs could still act as a seed to the SMBHs \cite{Bernal:2017nec}, and the constraints can be evaded if the initial perturbations are extremely non-Gaussian \cite{Nakama:2017xvq} although one then needs to evade the strong Planck constraints on dark matter isocurvature modes \cite{Tada_2015,Young_2015}.  For even broader power spectra the $\mu$--distortion constraints rule out an ever greater range of PBH masses, and for $\Delta=2$ they extend as far as the peak PTA constraint and thereby even rule out LIGO mass PBHs. Since such a wide peak in the primordial power spectrum provides the preferred PBH mass distribution width when fitting to LIGO data, it appears that the $\mu$-distortions may surprisingly provide a stronger constraint on models in which all LIGO black holes are PBHs than the PTA constraints. Of course this conclusion may also depend on the assumed shape of the power spectrum peak.

\begin{figure}[H]
\centering
\includegraphics[width=0.9\textwidth]{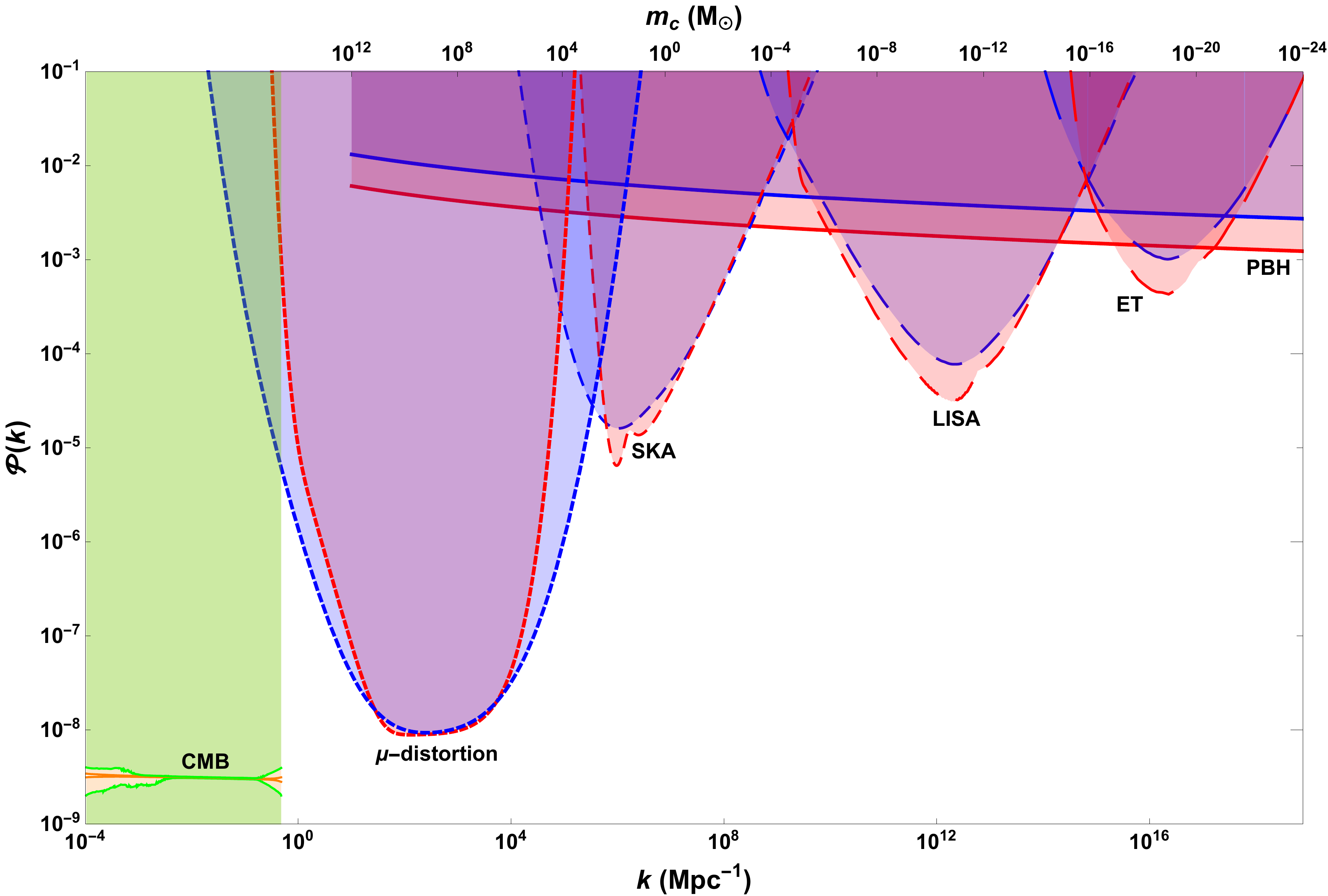}
\caption{Plot of the constraints on the power spectrum amplitude from PBH, gravitational wave background, and $\mu$-distortion sources, as well as the measured one and three-sigma values from the CMB. The PBH curves indicate the amplitude required to generate only a single PBH in the observable universe. The constraints for $\Delta=0.3$ (which are tighter for the PBH constraints, and narrower for the other constraints) are shown in red, and the constraints for $\Delta=1$ are shown in blue. The PBH constraints are shown with a solid line, and the ET, LISA, SKA, and $\mu$-distortion constraints are shown with longest to shortest dashes respectively.}
\label{fig:combinedConstraints_future}
\end{figure}

Future constraints from $\mu$-distortions and the gravitational wave background will significantly affect the PBH landscape. To examine the maximum extent of these future constraints, we calculate the PBH lines in the case that zero PBHs form in the observable universe. This is done using the method described in Ref.~\cite{Cole:2017gle}, particularly eq.~(7) of that paper, but with $\beta$ replaced with the $\Omega_{\PBH*}$ parameter used in this paper. For reasons summarised in Ref.~\cite{Cole:2017gle}, these extreme constraints might actually apply to the case of evaporated PBHs. Extremely tight constraints on $f_{\rm PBH}$ for $M_{\rm PBH}\gtrsim 10^{-6}\ \Msun$ are also possible if the majority of dark matter consists of ``standard'' WIMPs \cite{Lacki:2010zf,Eroshenko:2016yve,Boucenna:2017ghj,Adamek:2019gns,Bertone:2019vsk,Carr:2020xqk}. We show these constraints in fig.~\ref{fig:combinedConstraints_future}, as well as future $\mu$-distortion constraints from a detector like the Primordial Inflation Explorer (PIXIE) \cite{Chluba:2012_CMB}, and future gravitational wave background constraints from the Square Kilometre Array (SKA), the Laser Interferometer Space Antenna (LISA), and the Einstein Telescope (ET)\footnote{Note that free spectrum sensitivity curves, as were used to calculate the PTA constraints, are not available for the future detectors SKA, LISA, and ET, so instead we have used the sensitivity curves that are derived assuming a power-law for the gravitational wave frequency spectrum}. The SKA constraints are derived from the sensitivity curve calculated in Ref.~\cite{Moore:2014lga}, the LISA constraints are derived from the most optimistic sensitivity curve in fig.~1 of \cite{Bartolo:2016ami}, and the ET constraints are derived from fig.~13 of \cite{Maggiore:2019uih}.

It can be seen that the SKA constraints are so tight that a non-detection will indicate that no PBHs can exist in the LIGO range of masses, and hence that the LIGO merger events cannot possibly be explained with a primordial origin. Additionally, the combined effect of the $\mu$-distortion, SKA, LISA, and ET constraints removes the possibility of any PBHs existing over an extremely broad range of masses in the case of a non-detection, leaving only the space below $\sim10^{-22}\ \Msun$, and two small pockets at \mbox{$\sim 10^{-17}$--$10^{-14}\ \Msun$} and \mbox{$\sim 10^{-6}$--$10^{-3}\ \Msun$}.

\section{Conclusions}
\label{sec:conclusions}

We have made the first detailed analysis of how the PBH mass distribution shape and amplitude varies between three different techniques to calculate the primordial mass distribution: Press Schechter, traditional peaks theory and a newly developed peaks theory variation. We also consider two choices of the window function, a real-space top-hat and a modified Gaussian. We show that the amplitude of the primordial power spectrum only varies by $\mathcal{O}(10\%)$ for different choices, far smaller than may have been expected based on the large range of values of the power spectrum amplitude considered in the literature. A substantial variation remains depending on the shape of the peak in the primordial power spectrum, but this reflects a change in the physical theory rather than a change in methodology. The results are summarised in table~\ref{tab:fPBH_amplitudes} while fig.~\ref{fig:psiPlot_M-All} shows that the mass distribution shape hardly changes depending on the calculation technique. These differences, while not significant now, will be important for future data that probes the PBH mass distribution accurately, at which point an improvement of the TP method, such as the Young-Musso technique, should be used. We also show that the PBH mass distribution becomes broader as the power spectrum peak becomes broader, as highlighted in fig.~\ref{fig:psiPlot_Width-dependence}. In the limit of a narrow lognormal peak ($\Delta\lesssim0.3$) the mass distribution tends to a constant width which is set by critical collapse, making a peak of this width a well-motivated choice.

We have also calculated robust constraints on the primordial power spectrum from PBHs, taking into account the effects of critical collapse and the non-linear relation between $\zeta$ and $\delta$, as well as the choice of window function and the relation between the PBH mass scale and the peak power spectrum scale. This leads to tighter constraints that are shifted to different values of $k$ compared to those presented in Ref.~\cite{Carr:2020gox}. We show a summary of all of the key bounds on the amplitude of the primordial power spectrum in fig.~\ref{fig:combinedConstraints_current}. We stress that all the constraints must be recalculated when the shape of the primordial power spectrum peak is varied, and in the figure we choose $\Delta=0.3$ as a representative narrow peak and $\Delta=1$ as a broader peak. In both cases the PTA constraints (we use a recently improved data set from the NANOGrav collaboration) are almost identical to those from PBHs in the mass range that LIGO also probes. This interesting coincidence means that it is premature to rule out the possibility that LIGO detected PBHs that formed from large amplitude density perturbations during radiation domination, but if that is the case then there is a realistic hope that the PTA measurements will detect a stochastic background of gravitational waves in the near future and a dedicated analysis should be made. We note that the non-linear relation between $\zeta$ and $\delta$ weakens the PBH constraints by about a factor of 2, and had we not taken this into account (and normally it is not taken into account) we would have erroneously concluded that the PTA constraints do not come close to ruling out the formation of LIGO mass PBHs. However, we caution that if all BH binaries detected by LIGO were due to PBHs then the PBH mass distribution should be so broad ($\sigma_\psi\simeq0.8$ corresponding to $\Delta=2$) that the cosmic $\mu$-distortion constraints spread to relatively small masses and alternative shapes of the primordial power spectrum which are more ``top-hat''-like than the lognormal power spectrum studied here should be considered. 

In fig.~\ref{fig:combinedConstraints_future} we show constraints on the primordial power spectrum that could be achieved in the foreseeable future (assuming there is no detection) from a PIXIE-like experiment measuring $\mu$-distortions and searches for a stochastic background of gravitational waves. The gravitational wave constraints show SKA constraints on pulsar timings, plus LISA and ET constraints. The PBH constraints show the amplitude required to generate a single PBH within the observable universe, provided that they form from Gaussian-distributed perturbations entering the horizon during radiation domination. This shows that apart from two narrow mass ranges around $10^{-4}\ \Msun$ and $10^{-16}\ \Msun$, there will be no remaining window for unevaporated PBHs to exist today.

\section*{Acknowledgements}
CB thanks Qing-Guo Huang for correspondence, and we thank Eiichiro Komatsu for useful comments on a draft of this paper. AG is funded by a Royal Society Studentship by means of a Royal Society Enhancement Award. CB acknowledges support from the Science and Technology Facilities Council [grant number ST/T000473/1]. PC acknowledges support from the Science and Technology Facilities Council [grant number ST/N504452/1]. SY is supported by a Humboldt Research Fellowship for Postdoctoral researchers.

\appendix
\renewcommand\thefigure{A.\arabic{figure}}
\renewcommand{\theHfigure}{A.\arabic{figure}}
\setcounter{figure}{0}

\section{Ringing in the top-hat window function}\label{app:WF-TH}
\label{app:top-hat}

\begin{figure}[H]
\centering
\includegraphics[width=0.33\textwidth]{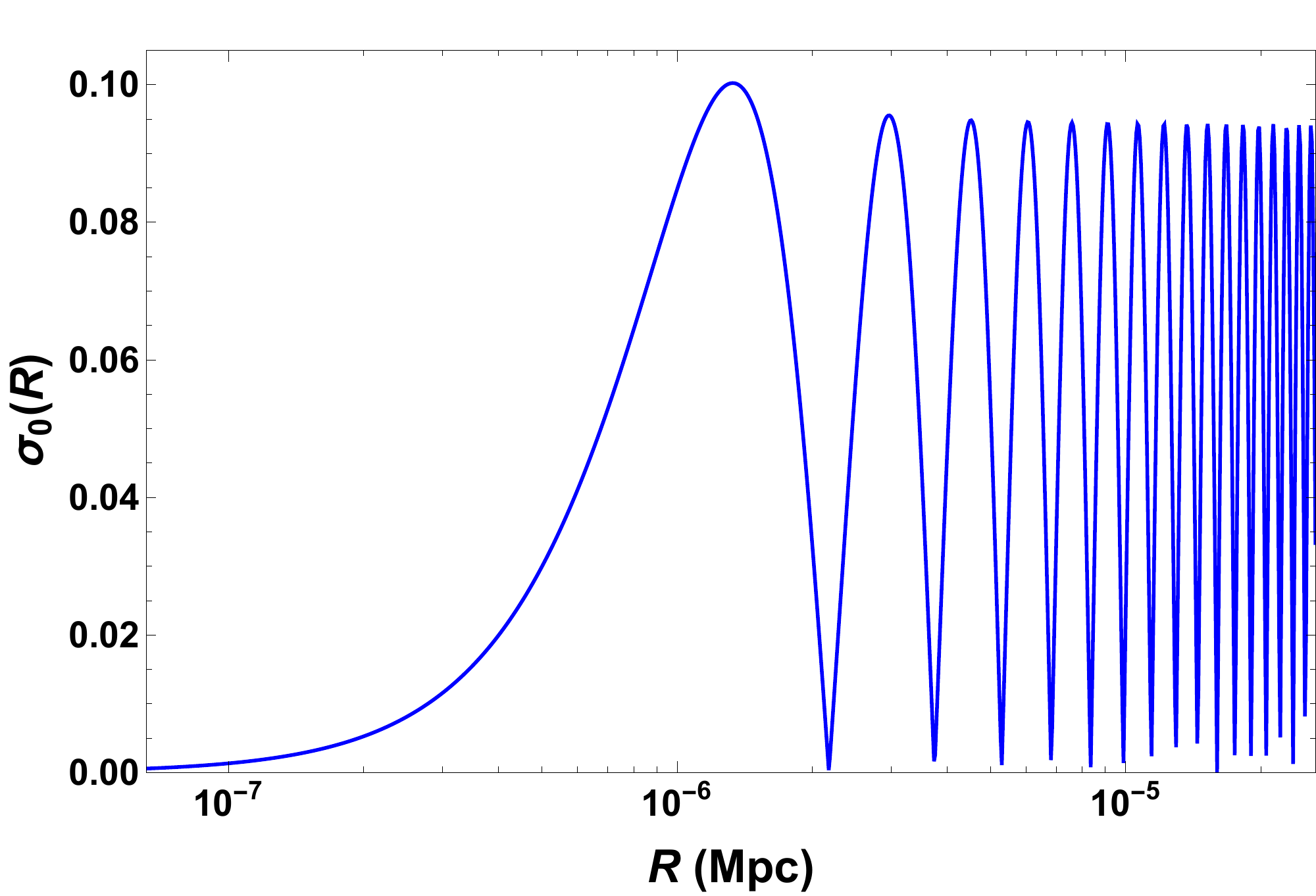}\includegraphics[width=0.33\textwidth]{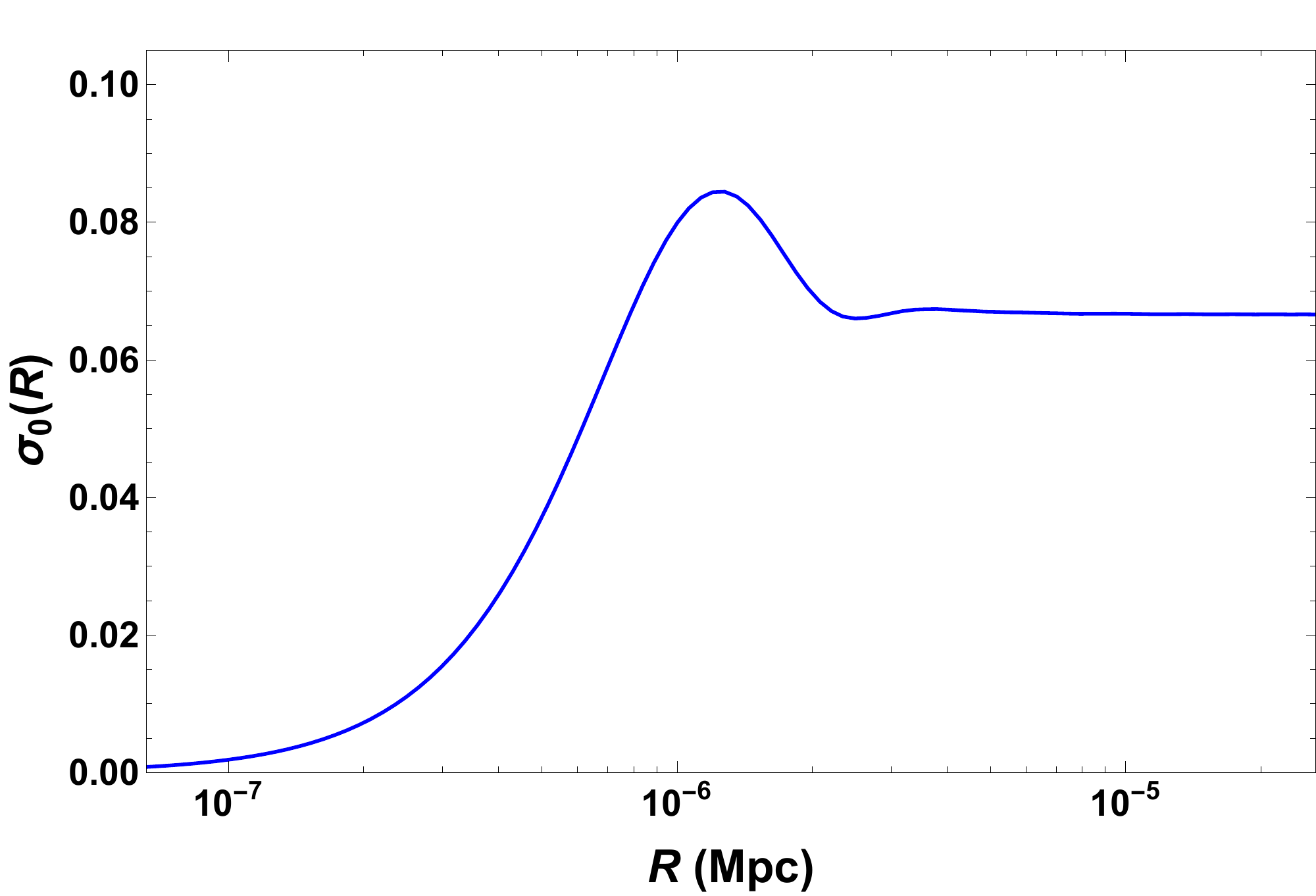}\includegraphics[width=0.33\textwidth]{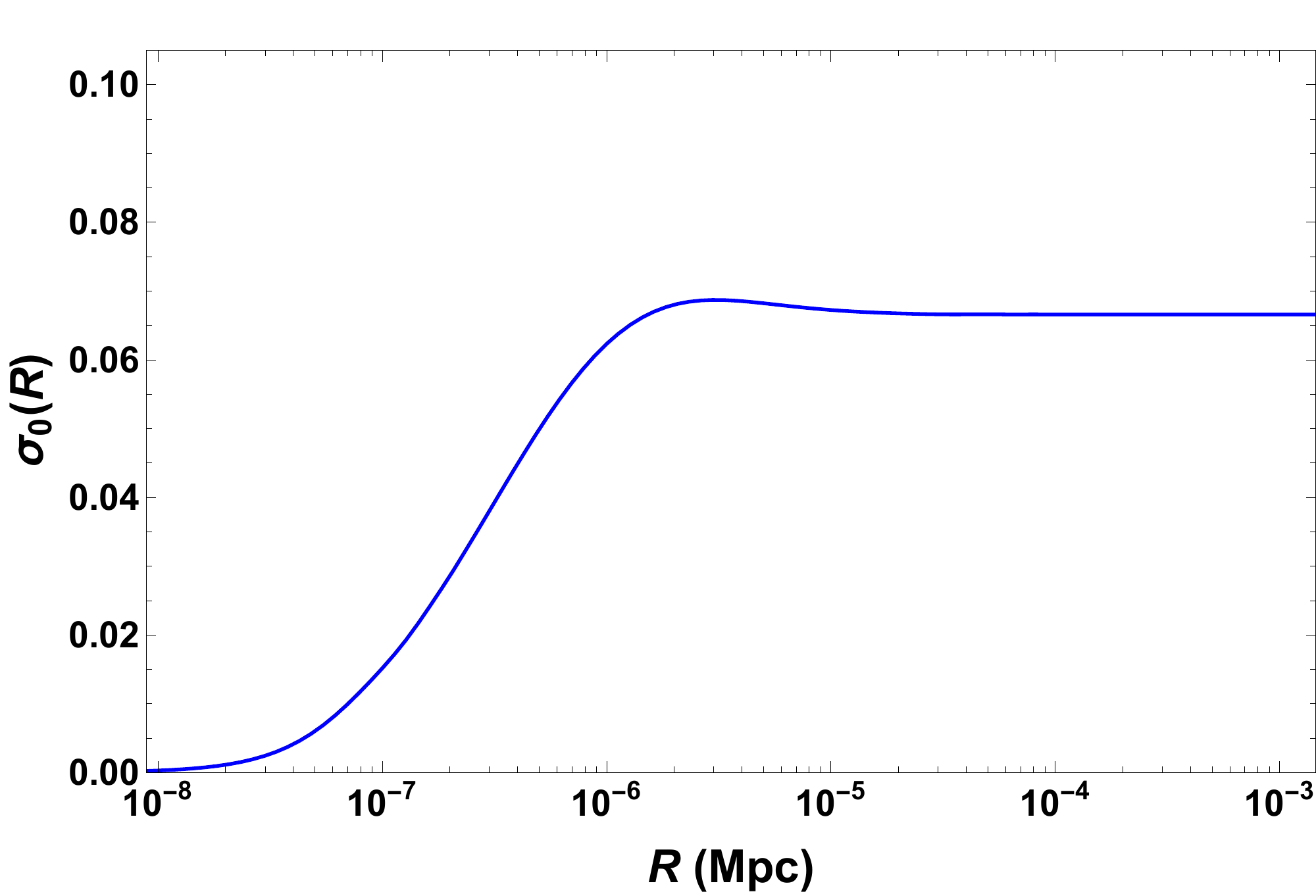}
\caption{Width parameter $\sigma_0(R)$ for a delta function power spectrum (left), and a lognormal peak with widths $\Delta=0.3$ (middle) and $\Delta=1$ (right). The ringing peaks visible in the delta case merge to a constant height as $\Delta$ increases.}
\label{fig:sigma0Plots}
\end{figure}

Here we explain our procedure to produce constraints when using a real-space top-hat window function, which corresponds to a rapidly oscillating window function in Fourier-space, with consequent convergence issues. The width parameter $\sigma_0(R)$ is shown in fig.~\ref{fig:sigma0Plots} for a delta function peak (left) and the lognormal widths $\Delta=0.3$ (middle) and $\Delta=1$ (right).

It can be seen that the oscillatory nature of the top-hat window function leads to a ringing effect in the width parameter $\sigma_0(R)$. For broader peaks in the power spectrum, this ringing effect merges into a constant height for large values of $R$. This leads to a divergent integral when evaluating eq.~\eqref{eq:Omega_PBH}, and so the mass distribution cannot be calculated with this window function without some form of adjustment. It is common to supress the large-$R$ constant effect using a transfer function, but this method is not compatible with other parts of our calculation (i.e.~\cite{Young:2019osy}). Therefore, we take an alternative approach, which is to adjust the calculation of $\sigma_n(R)$ in eq.~\eqref{eq:sigma_n} with a large-$k$ cutoff. This is placed at the point where the window function reaches its first trough, which is at $4.49/R$. This solves the divergence problem and removes the ringing/constant effect, but it must be noted that the window function is technically not a true top-hat any more.

\section{Observational constraints}\label{app:observations}

\subsection{Constraints due to spectral distortions of the CMB}
Spectral distortions of the energy spectrum of the CMB are able to constrain the primordial power spectrum on small scales. They quantify deviations from the black-body temperature distribution of the CMB, caused by energy injection and removal from the plasma in the early universe. A large boost in the primordial power spectrum at a particular scale or over a range of scales will lead to fluctuations in the density of the baryons and photons as a function of scale after reheating. This will mean that the photon distributions on different scales will be described by different blackbodies, and as those photons mix via Thomson scattering, a spectral distortion will be induced if Compton scattering, Double Compton scattering and Bremsstrahlung processes aren't efficient enough to bring them into equilibrium. So-called $y$-distortions quantify late-time processes and place constraints on larger modes $k<3\,{\rm Mpc^{-1}}$, whilst $\mu$-distortions quantify earlier energy injection and removal and hence constrain the smaller scales, up to $k\sim10^4\,{\rm Mpc^{-1}}$ which will be most interesting for PBH production. The final $\mu$-distortions induced by the scalar perturbations can be approximated by \cite{Chluba:2015bqa}
\begin{align}\label{eq:mu-dist}
\mu\approx&  \int_{k_{\rm min}}^\infty \frac{d k}{k} \mathcal{P}_\mathcal{R}(k) \, W_{\mu}(k),
\end{align}
with $k$-space window functions of the form
\begin{align}W_{\mu}(k) \approx& \, 2.27 \left[
	\exp\left(-\left[\frac{\hat{k}}{1360}\right]^2\Bigg/\left[1+\left[\frac{\hat{k}}{260}\right]^{0.3}+\frac{\hat{k}}{340}\right]\right) 
	- \exp\left(-\left[\frac{\hat{k}}{32}\right]^2\right)
	\right],
\end{align}
where $\hat{k}=k/1\,{\rm Mpc^{-1}}$ and $k_{\rm min}\simeq1\, {\rm Mpc^{-1}}$. Given a particular form for the power spectrum, this can be used to compute the total induced $\mu$ or $y$-distortion. Comparing this with observations then results in constraints on the primordial power spectrum. 

The Far-InfraRed Absolute Spectrophotometer (FIRAS) instrument on board the COsmic Background Explorer (COBE) satellite measured spectral distortions to be smaller than $\Delta\rho_\gamma/\rho_\gamma<6\times10^{-5}$ \cite{Fixsen:1996nj}, and a proposed future detector such as the Primordial Inflation Explorer (PIXIE) \cite{Kogut:2016_PIXIE}, or a more recent proposal \cite{Chluba:2019kpb} aims for constraints of $\Delta\rho_\gamma/\rho_\gamma<8\times10^{-9}$. To calculate the constraints on the amplitude of the power spectrum due to the COBE/FIRAS observations, we insert eq.~\eqref{eq:lognormal} into eq.~\eqref{eq:mu-dist} and set $\mu=9\times10^{-5}$ which is the 2-$\sigma$ constraint. We can then rearrange for $A$ and compute the integral over $k$, plotting the constraint on $A$ for each $k_p$. Our results for lognormal power spectra of widths $\Delta=0.3$ and $\Delta=1$ are shown in fig.~\ref{fig:combinedConstraints_current}. For complete clarity, the constraint on $\mathcal{P}_\mathcal{R}$ at a given $k$ represents the maximum amplitude $A$ for a lognormal power spectrum centred at $k=k_p$ so as not to induce $\mu$-distortions that would be in conflict with the COBE/FIRAS constraint of $\mu<9\times10^{-5}$.

\subsection{The stochastic gravitational wave background} 
\label{app:GW}

Here we summarise how the GW background can be calculated given a primordial power spectrum, adding more details to section \ref{sec:GW}. The contribution to the tensor power spectrum from the square of the scalar power spectrum is given by \cite{Kohri:2018awv,Espinosa:2018eve}
\begin{equation}
    \mathcal{P}_h(\tau, k) = 4 \int^\infty_0 dv
\int^{1+v}_
{|1-v|}du
\left(\frac{4v^2 - (1 + v^2 - u^2)^2}{4vu}\right)^2I^2
(v, u, k\tau)\mathcal{P}_\mathcal{R} (kv)\mathcal{P}_\mathcal{R} (ku),
\label{eq:Ph}
\end{equation}
where $u = |\textbf{k} - \tilde{\textbf{k}}|/k$, $v = \tilde{k}/k$ and $\tilde{k}$ is the wavelength corresponding to the scalar source. $I(v, u, k\tau)$ is a highly oscillatory function which contains the source information. We solve this integral numerically but note that it can be solved analytically in some regimes \cite{Pi:2020otn}. The observational quantity related to this power spectrum is the energy density of gravitational waves given by
\begin{equation}\label{eq:om}
    \Omega_{\rm GW}(\tau, k) = \frac{\rho_{\rm GW}(\tau, k)}{\rho_{\rm tot}(\tau)}=\frac{1}{24}\left(\frac{k}{aH}\right)^2\mathcal{P}_h(\tau, k).
\end{equation}

If we assume that the entire contribution to any stochastic background detection is from the tensor power spectrum in eq.~\eqref{eq:Ph}, then constraints on the stochastic background can be translated to constraints on the scalar power spectrum. This is a conservative constraint, as there may be other unresolved astrophysical contributions to the signal. If a detection is made, as opposed to an upper limit on the amplitude from non-detection, spectral information of the signal will be required to distinguish between the possible sources. To calculate the constraints on the primordial power spectrum, we first calculate $\Omega_{\rm GW}h^2$ today as a function of $k$ by inserting the lognormal power spectrum in eq.~\eqref{eq:lognormal} with given $k_p$ and $\Delta$ into eq.~\eqref{eq:Ph}, pulling out the amplitude $A$ which is the quantity that we aim to constrain. We perform this integral numerically once for each value of $\Delta$, and the results can be shifted post-integration for any value of $k_p$. 

\subsection{Updated NANOGrav dataset}
\label{app:ng}
\begin{figure}[H]
\centering
\includegraphics[width=0.8\textwidth]{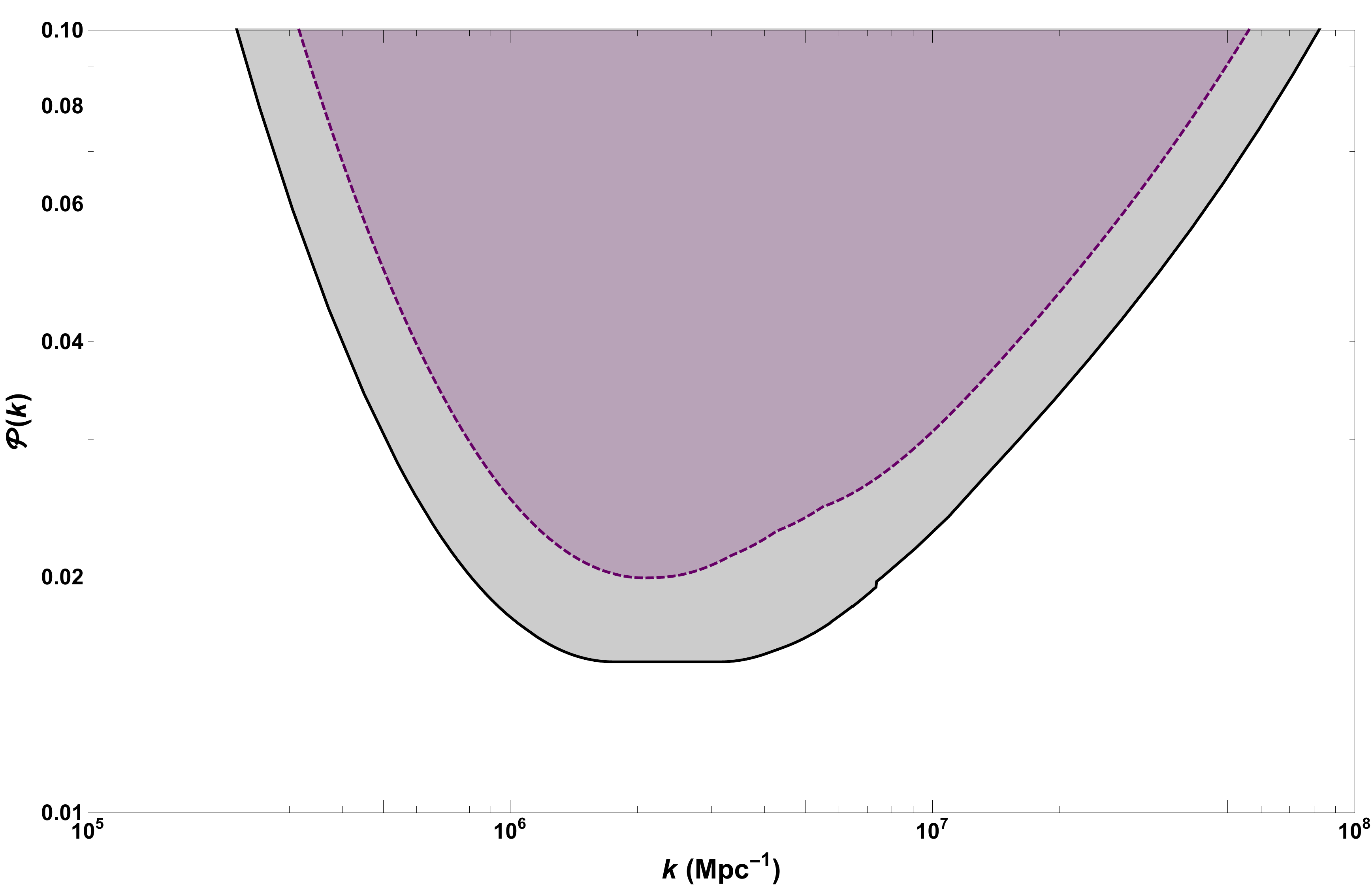}
\caption{Constraints on the amplitude of the primordial power spectrum due to NANOGrav pulsar timing array observations from the 9 year (purple, dashed) and 11 year (black, solid) datasets. For both datasets, constraints are for lognormal power spectra with width $\Delta=1$.}
\label{fig:ng119}
\end{figure}

The 11 year NANOGrav dataset \cite{Arzoumanian_2018} includes improved modelling of the solar system ephemeris which make the constraints on the stochastic gravitational wave background weaker than they would be with previous models of these effects. That makes the improvement on the primordial power spectrum constraints between the 11 year dataset and the 9 year dataset \cite{Arzoumanian_2016} not as large as one might hope based purely on the improved sensitivity. This solar system ephemeris modelling effect also applies to other pulsar timing array observations from, for example, EPTA \cite{Lentati_2015}. Therefore all of the constraints from these datasets need to be revised upwards by taking into account the better model for the solar system ephemeris. For this reason, we choose to just use the 11 year NANOGrav dataset alone, despite the fact that the EPTA dataset reaches lower frequencies, and as a guide to the improvement between datasets we show the constraint on the amplitude of the primordial power spectrum for a lognormal power spectrum with width $\Delta=1$ for both the 9 year and 11 year datasets in figure \ref{fig:ng119}.

\section{The non-linear relationship between $\zeta$ and $\delta$}
\label{app:nonLinearity}

In recent years, there has been a large amount of literature discussing the fact that, even if the curvature perturbation $\zeta$ is Gaussian, the density contrast will not be \cite{Young:2019yug,Yoo:2018kvb,Kawasaki:2019mbl,Kalaja:2019uju,DeLuca:2019qsy}, due to the non-linear relationship between the 2 parameters. In the super-horizon limit, the relationship between the 2 parameters can be calculated with a gradient-expansion approach. At first order in gradients, the full non-linear relationship, in polar coordinates and assuming spherical symmetry, is given by
\begin{equation} 
\delta_{NL}=\frac{\delta\rho}{\rho_b}(r,t) = - \frac{4(1+\omega)}{5+3\omega} \left(\frac{1}{aH}\right)^2 e^{-5\zeta(r)/2} \nabla^2 e^{\zeta(r)/2},
\label{eq:NLrelation}
\end{equation}
whilst the linear relation is
\begin{equation}
\delta_L=\frac{\delta\rho_l}{\rho_b} = -\frac{2(1+\omega)}{5+3\omega} \left(\frac{1}{aH}\right)^{2} \nabla^2\zeta. \label{eq:linearRelation}
\end{equation}
For simplicity, we will set the equation-of-state parameter $w=1/3$ from here on.

We can define a time-independent component of the density contrast,
\begin{equation}
\delta_\mathrm{TI}(\mathbf{x},R) = \left( \frac{1}{R\ aH} \right)^2 \delta_{NL},
\end{equation}
where $R$ is taken to be the scale of the perturbation.
The compaction function $C(\mathbf{x},R)$ is obtained by calculating the mass excess $\delta M$ within a sphere of radius $R$, and dividing by $R$, which corresponds to smoothing the time-independent component of the density contrast with a top-hat smoothing function,
\begin{equation}
C(\mathbf{x},R) = \frac{\delta M}{R} = \int \diffd^3y\ \delta_\mathrm{TI}(x-y)W(y,R). \label{eq:Compaction}
\end{equation}
Performing this integral gives an expression for the compaction function at the centre of spherically symmetric peaks:
\begin{equation}
C(\mathbf{x},R) =C_L-\frac{3}{8}C_L^2, \label{eq:NLquadratic}
\end{equation}
where $C_L$ is the expression one would obtain using the linear relation above,
\begin{equation}
C_L = \frac{2}{3}R \zeta'(R),
\end{equation}
where the prime denotes a derivative with respect to the smoothing scale $R$.

The rare, large-amplitude peaks from which PBHs form are well approximated by spherically-symmetric peaks \cite{Bardeen:1985tr}, and so the above equation can be used to relate relevant peaks in $C_L $ to peaks in the compaction $C$. We note that the compaction has a maximum value, $C_\mathrm{max} = 2/3$, corresponding to $C_L =4/3$. For higher values of $C_L$, the compaction decreases -- and perturbations of this type correspond to a case for which PBH formation has not been simulated. For this reason, only perturbations with $C_L <4/3$ are typically considered -- although in practice this has little effect on the PBH abundance since such large values of $C_L$ are exponentially suppressed.

If we then wish to calculate parameters related to the PBH abundance, we can simply replace the equation for the PBH mass, eq.~\eqref{eq:criticalScaling}, with a corresponding equation which relates the PBH mass to the linear, Gaussian component of the compaction instead
\begin{align}
m &= kM_H(C_L-\frac{3}{8}C_L^2-C_c)^\gamma.
\end{align}
In order to make an analytic estimate for how constraints on the power spectrum are affected by this non-linearity, we can make a simple assumption that all peaks which form PBHs are close to the critical amplitude (since the abundance of significantly larger peaks is exponentially suppressed). In this simple case, and assuming $C_c = 0.55$ (the case for the top-hat window function, see eq.~\eqref{eq:WF-TH}), the critical amplitude for the linear component of the compaction is $C_{c,L}\approx0.77$, i.e.~we can assume that peaks in the linear field need to have an amplitude 1.41 times larger than if we assumed a linear relation between $\zeta$ and $\delta$, as in eq.~\eqref{eq:linearRelation}. Therefore, the power spectrum (which is proportional to the variance of perturbations) should be approximately $1.41^2 = 1.98$ times greater. We can test this approximation by comparing the full calculation of the amplitude required to generate a fixed abundance $f_\PBH=2\times10^{-3}$ in the linear and non-linear cases. For the two lognormal power spectra considered in this paper, with widths $\Delta=1$ and 0.3, the approximation holds to the precision of two decimal places stated above. Although this validity may vary with the position of the peak, we assume it holds globally for the results shown in figs.~\ref{fig:combinedConstraints_current}~and~\ref{fig:combinedConstraints_future}.

For the top-hat window function, there is a relatively simple analytic relationship relating the compaction function to the curvature perturbation (which we assume to be Gaussian). However, we note that if one instead uses a Gaussian window function, as we have considered in this paper, there is no analytic solution, and accounting for the non-linearity becomes complicated. When looking at individual perturbations, it is trivial to show that the amplitude of the compaction (or ``compaction-like'') function calculated with both a top-hat or Gaussian window function is proportional to the amplitude of the perturbation. Therefore, we expect the non-linearities described above to have a similar effect on constraints on the power spectrum, whether a top-hat or Gaussian function is used.

\bibliographystyle{JHEP-edit} 
\bibliography{Power_spectrum_constraints_paper}{}

\end{document}